\documentclass[
	prd,
	preprint,
	aps,floats,floatfix,superscriptaddress,
	preprintnumbers,showpacs,nofootinbib,
]{revtex4-1}
\usepackage{
	amsfonts,amsmath,epsfig,amssymb,latexsym,
	eucal,color,graphicx,array,subfigure,bm,mathrsfs,
	ulem,epsfig,physics
}
\newcommand{\eq}[1]{\begin{align}#1\end{align}}
\newcommand{\pd}{\partial}

\begin{document}
%---------------------------------------------------------------------%

%---------------------------------------------------------------------%
\title{Determining parameters of a spherical black hole with a thin accretion disk by observing its shadow}
%---------------------------------------------------------------------%
%---------------------------------------------------------------------%
\author{Kenta Hioki}
\email{kenta.hioki@gmail.com}
\address{Sumitomo Mitsui Banking Corporation, 1-2, Marunouchi 1-chome, Chiyoda-ku, Tokyo 100-0005, Japan\footnote[1]{The statements expressed in this paper are those of the authors and do not represent the views of Sumitomo Mitsui Banking Corporation or its staff.}}
\author{Umpei Miyamoto}
\email{umpei@akita-pu.ac.jp}
\address{Research and Education Center for Comprehensive Science, Akita Prefectural University, Akita 015-0055, Japan}
%---------------------------------------------------------------------%
%---------------------------------------------------------------------%
\begin{abstract}
We revisit the classic system of a spherically symmetric black hole in general relativity ({\it i.e.}, a Schwarzschild black hole) surrounded by a geometrically thin accretion disk. Our purpose is to examine whether one can determine three parameters of this system ({\it i.e.}, black hole mass $M$, distance between the black hole and an observer $r_o$, inclination angle $i$) solely by observing the accretion disk and the black hole shadow. A point in our analysis is to allow $r_o$ to be finite, which is set to be infinite in most relevant studies. First, it is shown that one can determine the values of $(r_o/M, i)$, where $M/r_o$ is the so-called angular gravitational radius, from the size and shape of shadow. Then, it is shown that if one additionally knows the accretion rate $\dot{M}$ (respectively, mass $M$) by any independent theoretical or observational approach, one can determine the values of $(M, r_o, i)$ [respectively, $(\dot{M}, r_o, i)$] without degeneracy, in principle, from the value of flux at any point on the accretion disk.
\end{abstract}
%---------------------------------------------------------------------%
%---------------------------------------------------------------------%

\maketitle

%-------------------------------------------%
\section{Introduction}
\label{sec:intro}
%-------------------------------------------%

Black holes are quite interesting objects and provide us an ultimate test ground of strong gravitational fields. Therefore, it has been a mark to observe the shadow of a black hole forming when the light rays emitted from the ambient material are bent by the gravitational field of black hole. 

Recently, the image of M87, which has been a black hole candidate, was indeed captured by an Earth-size very long baseline interferometer~\cite{EventHorizonTelescope:2019dse, EventHorizonTelescope:2019uob, EventHorizonTelescope:2019jan, EventHorizonTelescope:2019ths, EventHorizonTelescope:2019ggy, EventHorizonTelescope:2021bee, EventHorizonTelescope:2021srq}, and that of ${\rm Sgr\; A^\ast}$ was captured, too~\cite{EventHorizonTelescope:2022xnr, EventHorizonTelescope:2022vjs, EventHorizonTelescope:2022wok, EventHorizonTelescope:2022exc, EventHorizonTelescope:2022urf, EventHorizonTelescope:2022xqj}. One method to confirm the existence of black holes by capturing the image can be said to be established (but see also Ref.~\cite{Miyoshi:2022eor}). While the images of black hole systems have been observed, the shadow itself, namely, the dim part, has not been observed yet, and further improvements of observational equipment are said to be needed~\cite{EventHorizonTelescope:2019ths}.

One might say that the study of black hole imaging was started with the derivation of shadow contour or, as we now call it, apparent shape.
The apparent shape of a Schwarzschild black hole was first derived in Ref.~\cite{darwin1959gravity}, and that of a Kerr black hole was done in Ref.~\cite{Bardeen:1973xx}.
Note that Ref.~\cite{Synge:1966okc} is the second reference of Ref.~\cite{darwin1959gravity}. These works should be said to establish the basis of the shadow theory. While they calculated the apparent shapes of simple ``bare'' black holes, namely, did not take into account accretion disks around the black holes, the role of photon sphere was revealed, which plays a central role even in the shadow of black hole with the accretion disk.

The image of the Schwarzschild black hole with the accretion disk was derived in Ref.~\cite{Luminet:1979nyg}. The image of the Kerr black hole with the accretion disk was done in Refs.~\cite{Falcke:1999pj, Takahashi:2004xh} for several fixed values of the parameter. The numerical study of shadow using the models that can be thought to mimic real situations also made remarkable progress~\cite{James:2015yla, Cunha:2019hzj, Dokuchaev:2020wqk, Chael:2021rjo}. The apparent shapes and images of various black hole solutions have also been obtained~\cite{Hioki:2008zw, Bambi:2010hf, Amarilla:2010zq, Amarilla:2013sj, Wei:2013kza, Papnoi:2014aaa, Wei:2015dua, Singh:2017vfr, Stuchlik:2019uvf}.

So far, many researchers have considered what shadows of black holes look like and how to extract physical information such as the angular momentum of the black hole by observing shadows. Nevertheless, what we would like to insist in this paper is that there is still one direction to improve the decidability of physical parameters of black holes from the shadow. In particular, it remains to be examined whether it is possible or not to extract information such as the angular momentum of the black hole by observing the shadow. 

One method to examine the above possibility is to investigate whether or not a map from a parameter space to an image library is an injection~\cite{Hioki:2009na}. One of the present authors (K.H.) and his collaborator, based on this idea, showed that the map from the parameter space to the apparent-shape library for a bare Kerr black hole is indeed a bijection, which means that the angular momentum (per mass squared) and inclination angle of the Kerr black hole can be determined by observing its apparent shape. Here, the apparent-shape library is defined as the set of all possible apparent shapes that can be generated  in the given gravitational theory and model.

Subsequently, the observables of black hole shadows which characterize the shadow were improved~\cite{Abdujabbarov:2015xqa}, and geometric analysis of the apparent shapes was developed~\cite{Wei:2019pjf}. Using the improved observables and actual data, an attempt was made to identify the black hole solution describing M87 and to put restrictions on its physical parameters~\cite{EventHorizonTelescope:2021dqv}.

Is it possible to i) confirm that it is a black hole, ii) identify the black hole solution, and iii) determine its physical quantities only by observing the shadow image of a black hole candidate object? To answer this question, we presumably have to conduct laborious research. Namely, we have to prepare a huge amount of models of relativistic objects and accretion disks and construct their image libraries by allowing their parameters to change. Then, we also have to investigate whether the map is injective (or invertible). If it is not injective, it means that there exists an image corresponding to different models and parameter settings, and it will be impossible to determine the model based on shadow observation alone. These efforts are currently in progress, and further studies are required.

%-------------------------------------------%
\begin{figure}[bt]
	\includegraphics[width=7cm]{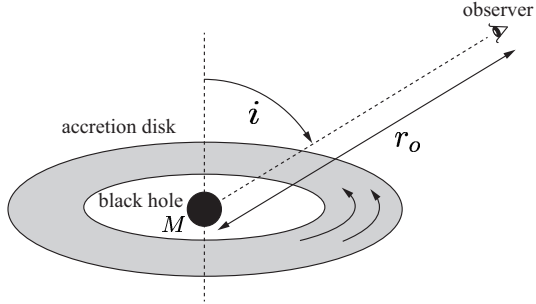}
\caption{A schematic picture showing three parameters $(M,r_o,i)$ of the system considered in this paper. $M \; (>0)$ is the mass of a Schwarzschild black hole surrounded by an infinitely thin rotating accretion disk. An observer is located at distance $r_o \in [20M, +\infty)$ from the center of the black hole in the geometrical units ($c=G=1$). Inclination angle $i \in [0^\circ , 90^\circ )$ is defined as the angle between the rotation axis of the accretion disk and observer. The distance of the inner and outer edges of accretion disk from the center of black hole are $6M$ ({\it i.e.,} the innermost stable circular orbit) and $20M$, respectively. The results in this paper are independent of the position of outer edge.}
\label{fg:situation}
\end{figure}
%-------------------------------------------%

In this paper, as a first step to answer the above questions, we reinvestigate the classic model of the Schwarzschild black hole with the thin accretion disk considered in Ref.~\cite{Luminet:1979nyg}, from a different point of view. Namely, we allow the distance between the black hole and the observer to be {\it finite} rather than infinite. The reason is that we expect a large number of black hole candidates to be observed in the future, all of which are at different distances from us, and we would like to present a method taking into account finite-distance effects precisely. The three parameters characterizing our simple system are shown in Fig.~\ref{fg:situation}. Note that a few methods to take into account the finite-distance effects in shadow observation were proposed~\cite{Grenzebach:2014fha, Abdolrahimi:2015rua}.

As the results, we will see that one can determine only the angular gravitational radius $M/r_o$ and inclination angle $i$ from the observation of apparent shape of the shadow. We will see, however, that one can determine the mass $M$, distance $r_o$, and inclination angle $i$ separately if the value of bolometric energy flux at any point on the accretion disk is observed and if the value of accretion rate $\dot{M}$ is known from any independent theoretical or observational approach.

Reference~\cite{James:2015yla} reveals what a Kerr black hole surrounded by an accretion disk looks like from a moving observer at finite distance, the results of which are used in the Hollywood film {\it Interstellar}.~\footnote{We thank the anonymous referee for informing us of this highly relevant paper.} While such a setup in Ref.~\cite{James:2015yla} is more general than ours in the present paper, we would like to stress that our main aim is not revealing what the black hole surrounded by the accretion disk looks like for the finite-distant observer but examining whether one can extract physical parameters of the black hole from a two-dimensional image. As you will see soon, such a decidability of parameters from the image is not so trivial, which is the reason why we restrict ourselves to the simplest case of Schwarzschild black hole with the infinitely thin accretion disk in this paper. Thanks to such a simplification, we could manage to complete the analysis in a semianalytic way and succeed in showing the determinability explicitly. We will discuss, however, the decidability of parameters from an image for a finite-distance Kerr black hole in the next paper.

The organization of this paper is as follows. In Sec.~\ref{sec:geodesics}, we briefly review the behaviors of null and timelike geodesics around a Schwarzschild black hole, which correspond to the motion of the photons emitted from the accretion disk and massive particles in the accretion disk, respectively. In Sec.~\ref{sec:setup}, we describe the system in a correct manner, how to define a two-dimensional image from the null geodesics, and how to determine the system's parameters from observation. In Sec.~\ref{sec:shape}, we present the results on the apparent shape of the black hole obtained by using the formulation prepared in previous sections. It is shown that one can determine the values of $(r_o/M, i)$ by observing the size and shape of shadow. In Sec.~\ref{sec:flux}, it is shown that further observational information, {\it i.e.}, the flux at any point on the accretion disk and the mass accretion rate, makes it possible to determine the values of $(M, r_o,  i)$ without degeneracy. In Sec.~\ref{sec:discussion}, a comparison of our results with those of related papers will be discussed. We summarize our analysis and mention future prospects in the final section. We use the geometrical units, in which $c=G=1$, throughout this paper.

%-------------------------------------------%
\section{Geodesics in Schwarzschild spacetime}
\label{sec:geodesics}
%-------------------------------------------%
The line element in the Schwarzschild black hole is
\begin{eqnarray}
	g_{\mu\nu}(x) dx^\mu dx^\nu
	=
	-f(r) dt^2 + \frac{1}{f(r)} dr^2 + r^2 (d\theta^2 + \sin^2 \theta d\phi^2),
	\;\;\;
	f(r) := 1 - \frac{2 M}{r},
	\label{eq:line_element}
\end{eqnarray}
where $x^\mu = (t,r,\theta,\phi) \; (\mu,\nu = 0,1,2,3)$ and $M \; (>0)$ is the mass of black hole. The geodesic equations for a massless particle and a massive particle are obtained as the Euler-Lagrange equation
\begin{eqnarray}
	\frac{d}{d\lambda}\left(\frac{\pd {\cal L}}{\pd \dot{x}^\mu}\right) - \frac{\pd {\cal L}}{\pd x^\mu}=0,
\label{eq:EL}
\end{eqnarray}
with a respective suitable Lagrangian ${\cal L}$. Here, the dot represents the derivative with respect to $\lambda$, parametrizing the geodesic $x^\mu (\lambda)$.

%-------------------------------------------%
\subsection{Null geodesics}
\label{sec:null}
%-------------------------------------------%

The Lagrangian for a massless particle on equatorial plane $\theta= \pi /2$ is
\begin{eqnarray}
	{\cal L}
	=
	\frac12 g_{\mu\nu}\frac{dx^\mu}{d\lambda}\frac{dx^\nu}{d\lambda}
	=
	- \frac12 f(r) \dot{t}^2 + \frac12 f(r)^{-1}\dot{r}^2 + \frac12 r^2 \dot{\phi}^2.
\label{eq:Lag1}
\end{eqnarray}
From $t$ and $\phi$ components of the Euler-Lagrange equation~\eqref{eq:EL} with Lagrangian~\eqref{eq:Lag1}, we obtain
\begin{eqnarray}
p_t
	=
	- f(r)\dot{t}= -E,
\;\;\;
	p_\phi
	=
	r^2 \dot{\phi} 
	= L,
\label{eq:tphidot}
\end{eqnarray}
where $p_\mu := \frac{\pd {\cal L}}{\pd \dot{x}^\mu}$, and $E$ and $L$ are integration constants. In the case of a massless particle, $\lambda$ is an affine parameter. Combining Eq.~\eqref{eq:tphidot} with null condition ${\cal L}=0$, we obtain 
\begin{eqnarray}
	\left(\frac{dr}{d\lambda'}\right)^2 &=& 1-\frac{b^2 f(r)}{r^2},
\label{eq:drdlambda'}
\end{eqnarray}
where $\lambda' := E \lambda $, and $b:=L /E$ is an impact parameter. Substituting Eqs.~\eqref{eq:tphidot} and \eqref{eq:drdlambda'} into chain rule $\frac{dr}{d\phi} = \frac{dr}{d \lambda'} \frac{d\lambda'}{d\phi}$, we obtain the equation for trajectory $r=r(\phi)$ in a potential form,
\eq{
	\left(\frac{1}{r^2}\frac{dr}{d\phi}\right)^2 + V(r) = \frac{1}{b^2},
\;\;\;
	V(r) := \frac{f(r)}{r^2}.
\label{eq:energy_cons}
}

\begin{figure}[htb]
	\includegraphics[height=4cm]{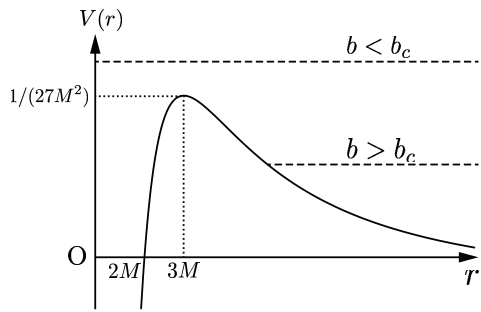}
	\caption{\footnotesize{A schematic picture of effective potential $V(r)$ for a massless particle around the Schwarzschild black hole [see Eq.~\eqref{eq:energy_cons}]. Two horizontal dashed lines represent lines of $1/b^2$. The lower line ($b>b_c:=3\sqrt{3}M$) and upper line ($b<b_c$) correspond to the null geodesic of the first kind and with an imaginary eccentricity, respectively.}}
\label{fg:Veff1}
\end{figure}

As shown in Fig.~\ref{fg:Veff1}, effective potential $V(r)$ has a maximum $V(3M)=1/(27M^2) =: 1/b_c^2$, where $r=3M$ is the photon sphere. According to Ref.~\cite{Chandrasekhar:1985kt}, we call a null geodesic with impact parameter $b$ larger than $b_c=3\sqrt{3} M$ that of {\it the first kind} and call a null geodesic with impact parameter $b$ smaller than $b_c$ that with {\it an imaginary eccentricity}. These two kinds of geodesics, both of which play central roles in our analysis, can reach sufficiently far region (namely, an observer), provided they are emitted outward from a point with $r>3M$.

%-------------------------------------------%
%-------------------------------------------%

If one considers a geodesic moving outward ($\frac{dr}{d\phi} > 0$) from $r=r_1$ to $r=r_2 \; (r_1<r_2)$, the change of $\phi$ during such a motion is obtained by integrating Eq.~\eqref{eq:energy_cons},
\eq{
	\phi (r_2) - \phi (r_1)
	=
	\int_{r_1}^{r_2} \frac{b}{ r \sqrt{ r^2- b^2 f(r) }} dr.
\label{eq:phi_int}
}
As we will see in the next section, the captured image of the accretion disk is made by both the geodesics of the first kind and with imaginary eccentricity. For both kinds of geodesics, the integral in Eq.~\eqref{eq:phi_int} can be written down in terms of the incomplete elliptic integrals of the first kind.

%-------------------------------------------%
\subsection{Timelike geodesics}
\label{sec:timelike}
%-------------------------------------------%
The Lagrangian for a massive particle moving on equatorial plane $\theta = \pi /2$ is
\eq{
	{\cal L}_m
	=
	\frac12 m g_{\mu\nu}\frac{dx^\mu}{d\tau}\frac{dx^\nu}{d \tau}
	=
	- \frac12 m f(r) \dot{t}^2 
	+ \frac12 m f(r)^{-1} \dot{r}^2 
	+ \frac12 m r^2 \dot{\phi}^2,	
\label{eq:Lag2}
}
where $m \; (>0)$ is the mass of particle and the dot represents the derivative with respect to particle's proper time $\tau$ here. From $t$ and $\phi$ components of Euler-Lagrange equation~\eqref{eq:EL} with Lagrangian \eqref{eq:Lag2}, we obtain
\eq{
	m f(r) \dot{t} = E_m,
\;\;\;
	m r^2 \dot{\phi} = L_m,
\label{eq:tphidotm}
}
where $E_m$ and $L_m$ are integration constants. Combining Eq.~\eqref{eq:tphidotm} with normalization condition $g_{\mu \nu} \dot{x}^\mu \dot{x}^\nu = -1$, we obtain the equation of radial motion in a potential form, 
\eq{
	\qty( \frac{dr}{d\tau} )^2
	+
	U(r) = \frac{E_m^2}{m^2},
\;\;\;
	U(r) := \qty(1+\frac{L_m^2}{m^2 r^2}) f(r).
\label{eq:energy_cons2}
}

\begin{figure}[htb]
	\includegraphics[height=4cm]{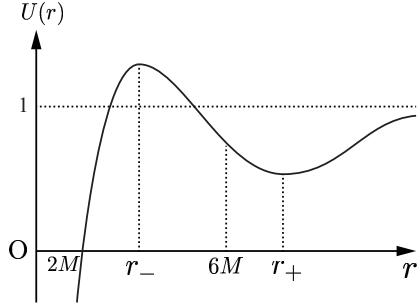}
	\caption{\footnotesize{A schematic picture of effective potential $U(r)$ for a massive particle around the Schwarzschild black hole, which depends on angular momentum of particle $L_m$ [see Eq.~\eqref{eq:energy_cons2}]. As in this picture, the potential has critical points at $r=r_-$ and $r=r_+ \; (3M< r_- < 6M < r_+)$ for $L_m > L_c := 2\sqrt{3}mM$. In the limit of $L_m \to L_c + 0$, the both critical points approach $6M$, which corresponds to the position of the ISCO.}}
	\label{fg:Veff2}
\end{figure}

The behavior of effective potential $U(r)$ depends on the angular momentum $L_m$. For a particle with a relatively large angular momentum, characterized by $L_m > L_c := 2\sqrt{3} mM$, $U(r)$ has critical points at $ r = r_\pm \;  (3M< r_- < 6M < r_+)$, where $r = r_- $ and $r = r_+ $ correspond to unstable and stable circular orbits, respectively (see Fig.~\ref{fg:Veff2}). For $L_m = L_c$, $r_-$ and $r_+$ coincide to be $6M$, which is the radius of {\it innermost stable circular orbit} (ISCO). For $L_m < L_c$, both critical points do not exist.

%-------------------------------------------%
%-------------------------------------------%
The angular velocity of a particle $\varOmega$ is given by
\eq{
	\varOmega
	=
	\frac{\dot{\phi}}{\dot{t}}
	=
	\frac{L_m f(r)}{ E_m r^2},
\label{eq:Omega1}
}
where we have used Eq.~\eqref{eq:tphidotm} in the last equality. Now, let us consider a perfectly circular orbit or a Keplerian motion. The radius of such an orbit (namely, $r_+$) is determined by two conditions of $U(r)=E_m^2/m^2$ and $U'(r)=0$. Eliminating from $E_m$ and $L_m$ from the right-hand side of Eq.~\eqref{eq:Omega1} using these two conditions, we can represent the angular velocity of the Keplerian motion in terms of its radius,
\eq{
	\varOmega = \sqrt{ \frac{M}{r^3} }.
\label{eq:Omega2}
}
It is well known that this result coincides with one obtained from a Newtonian argument, namely, the balance between gravitational force $Mm/r^2$  and centrifugal one $m r \varOmega^2$.

We will assume all particles in the accretion disk to be in the Keplerian motion. Then, from Eq.~\eqref{eq:Omega2}, closer to the center, the angular velocity is greater; namely, the rotation is differential. Therefore, the inner material exerts a torque on the outer material in the direction of rotation through viscous stresses. Such viscous stresses transport angular momentum outward through the disk. The material that loses angular momentum spirals gradually inward until $r=6M$ (ISCO) and finally falls into the black hole. The viscous stresses, working against the differential rotation, also play the role of heating the disk to cause it to emit a large amount of flux~\cite{review}.

For a black hole shadow to form, a light source is required. When there is a sufficient number of light sources in every direction, the photon sphere of black hole, which is located at $r=3M$ in the Schwarzschild case, determines the shape of the black hole shadow. On the other hand, when there is only a thin accretion disk as a light source, which is the case of the present analysis, the photon sphere does not play any special role, provided only the direct image of black hole is concerned. In such a case, the shadow boundary is formed by the light rays from the inner edge of accretion disk, located at the ISCO. Note that the position of the ISCO is determined only by the mass of the central black hole, $r=6M$, and irrelevant of the detail of materials on the accretion disk.

%-------------------------------------------%
\section{Setup}
\label{sec:setup}
%-------------------------------------------%

Although the system of the black hole, accretion disk, and observer we investigate in this paper is quite simple as shown in Fig.~\ref{fg:situation}, let us describe it little more precisely here. Then, let us explain how to define two-dimensional image of the subject ({\it i.e.}, the accretion disk) from the information encoded in the null rays emitted from it. This is indispensable because the shape and size of the image may depend on the definition of the two-dimensional image, in particular, when the subject is close to the observer. Finally, we will define a map from a parameter space to the set of images, which is necessary for later investigation.

%-------------------------------------------%
\subsection{Black hole surrounded by an accretion disk and observer}
\label{sec:position}
%-------------------------------------------%

\begin{figure}[thb]
	\begin{center}
		\tabcolsep=2mm
		\begin{tabular}{cc}
			\includegraphics[height=45mm]{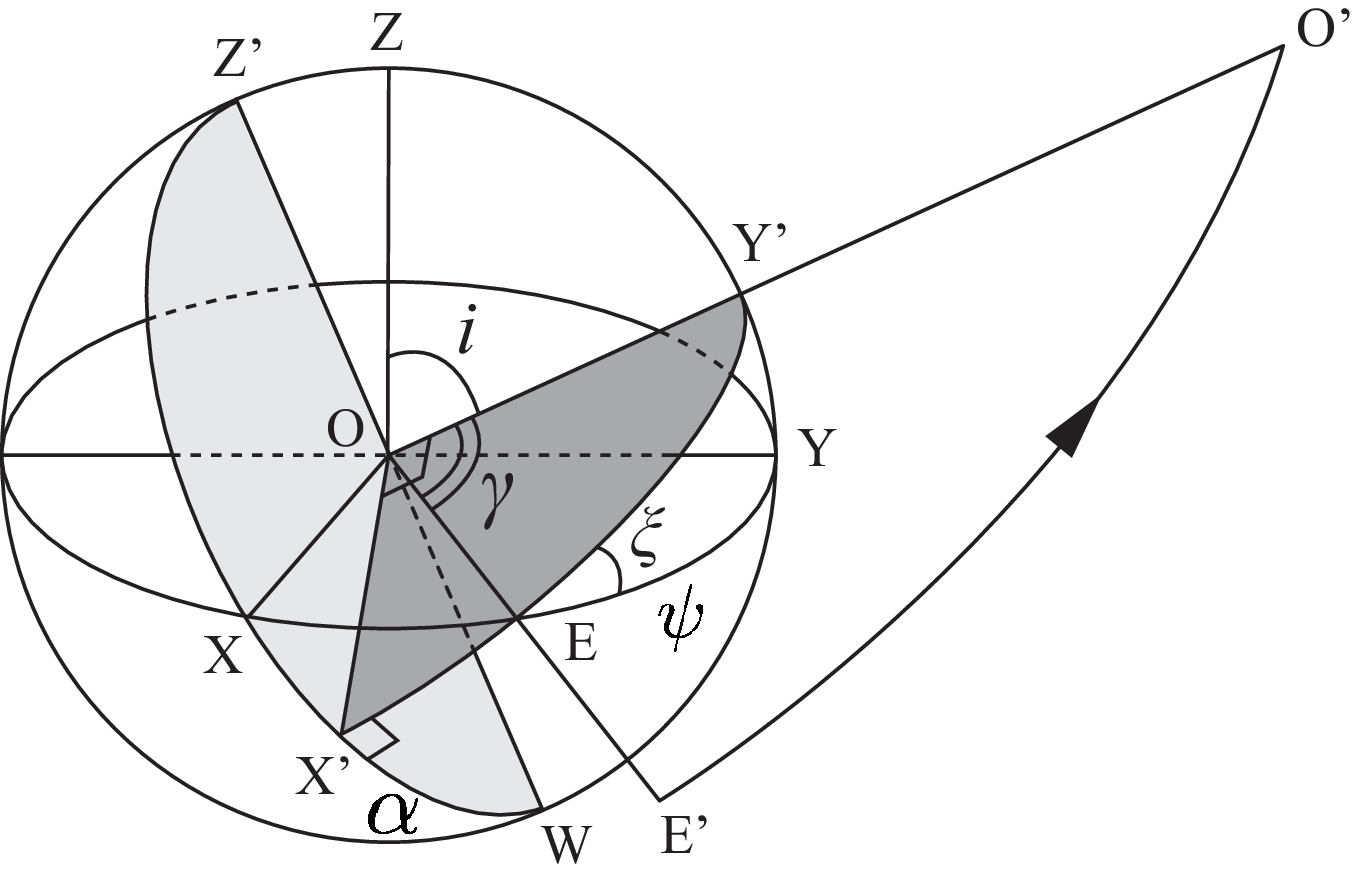} &
			\includegraphics[height=50mm]{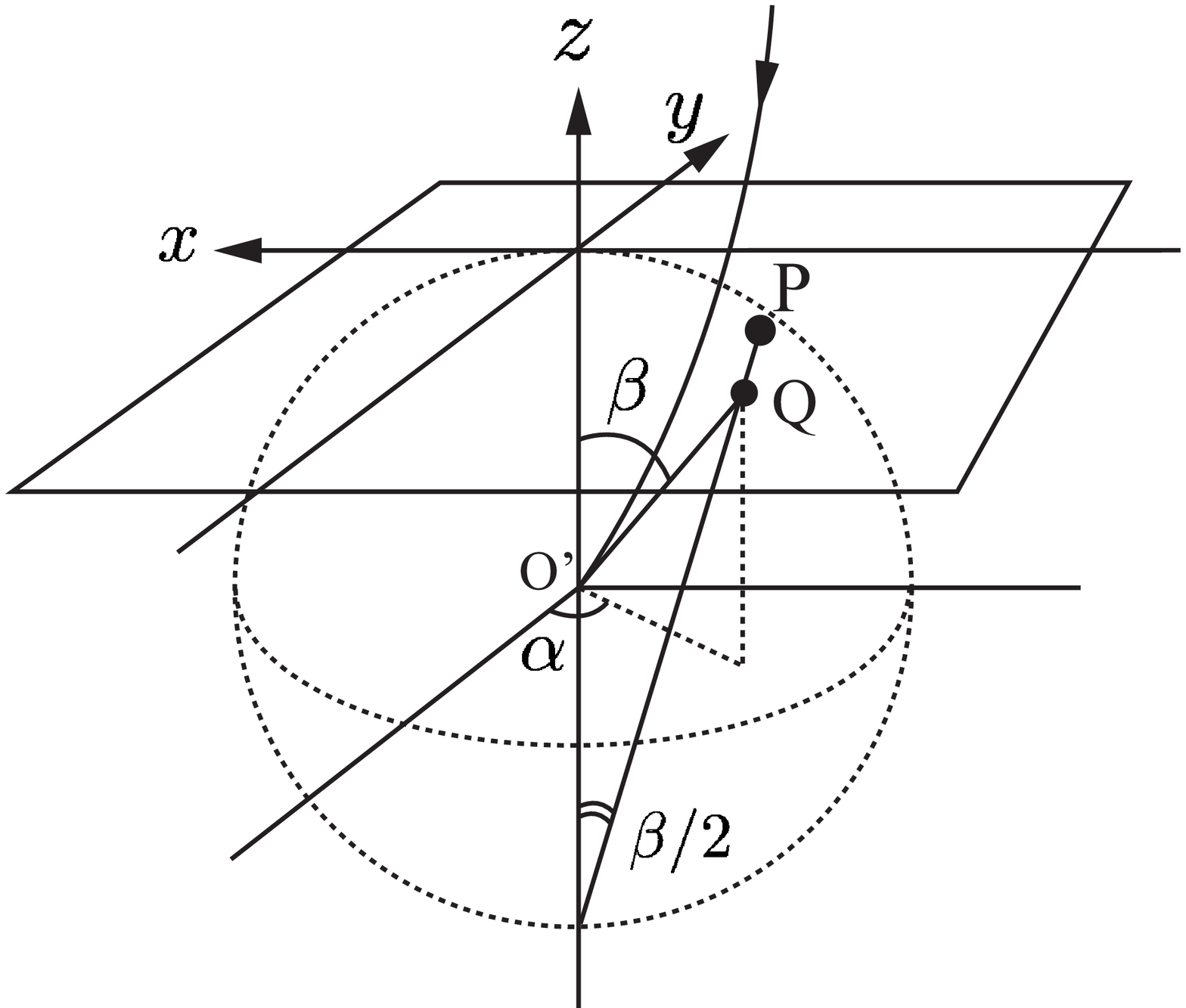} \\
			(a) & (b) \\
		\end{tabular}
		\caption{(a) A schematic picture showing a unit sphere of which center ${\rm O}$ coincides with that of the Schwarzschild black hole. The observer is located at ${\rm O}'$, which is on the $X=0$ plane. The accretion disk, which is a rotating annulus, is on the $Z=0$ plane. A null ray emitted at ${\rm E}'$ by a massive particle in the accretion disk reaches the observer along null ray ${\rm E'O'}$. See the text for the detail. (b) A schematic picture showing a unit celestial sphere for the observer ${\rm O'}$. The $+z$ direction is in the direction of ${\rm O}$. $(\alpha,\beta)$ is a celestial coordinate system, which is used to specify the incident angle of the null ray into the observer. ${\rm Q}$ is the point where the tangent of null ray at ${\rm O}'$ crosses the unit sphere. ${\rm P}$ is the two-dimensional image made by the null ray.}
\label{fg:angles}
	\end{center}
\end{figure}

The positions of the black hole, accretion disk, and observer are schematically shown in Fig.~\ref{fg:angles}(a). In this figure, a unit sphere of which center ${\rm O}$ coincides with that of the Schwarzschild black hole is drawn. Here, we introduce Schwarzschild coordinates $(t,r,\bar{\theta},\bar{\phi})$ by letting its origin $r=0$ coincide with ${\rm O}$ in Fig.~\ref{fg:angles}(a). While we identify coordinates $(t, r)$ with those in Eq.~\eqref{eq:line_element}, let us stress that coordinates $(\bar{\theta}, {\bar \phi})$, which are related to $(X,Y,Z)$ in the standard way, are {\it independent} of $(\theta,\phi)$ in Eq.~\eqref{eq:line_element} at this point.

In terms of the Schwarzschild coordinates introduced above, the accretion disk, which is an optically thick rotating annulus, is on the equatorial plane $\bar{\theta}= \pi /2$, and its inner and outer edges are at $r = 6M$ (ISCO) and $r=20M$, respectively. The observer rests at ${\rm O}'$, the position of which is specified by $(r,\bar{\theta},\bar{\phi}) = ( r_o, i, \pi /2 )$. Here, $r_o$ and $i$ are in the range of $20M \leq r_o <\infty$ and $0 \leq i < \pi /2$, respectively. A photon emitted by a massive particle in the accretion disk at ${\rm E}'$, the coordinates of which are $(r,\bar{\theta},\bar{\phi}) = (r_e, \pi /2, \pi /2 - \psi )$, reaches the observer along the null geodesic ${\rm E'O'}$. Needless to say, $r_e$ and $\psi$ are assumed to be in the range of $6M \leq r_e \leq 20M$ and $0 \leq \psi < 2 \pi$, respectively. We stress that the position of the outer edge of the accretion disk, $r=20M$, has no special meaning in the sense that the conclusion in this paper does not depend on this assumption.

%-------------------------------------------%
%-------------------------------------------%

As mentioned before, all particles in the accretion disk are assumed to be in the Keplerian motion ({\it i.e.}, a perfectly circular orbit). We also assume that every particle in the accretion disk isotropically radiates null rays in all directions. Among such null rays, only ones with appropriate impact parameter $b$ can reach the observer to generate the image, and the remaining null rays either fall into the black hole or escape to the asymptotic region. The shadow of the black hole is defined as the region in the two-dimensional image where no null ray reaches and which is surrounded by the image of the accretion disk. The apparent shapes of the black hole and accretion disk are defined by a boundary of a black hole shadow and a boundary of an accretion disk image, respectively.

For simplicity, we consider only the primary image, which is generated by primary rays, in this paper. Namely, we do not consider the secondary and higher images, generated by the photons that have circled the black hole once or more before reaching the observer.

%-------------------------------------------%
\subsection{How to define two-dimensional image}
\label{sec:image}
%-------------------------------------------%

We have to relate the information about the null rays to the two-dimensional image of the accretion disk in a way that works even when $r_o$ is finite. For such a method, in this paper, we employ a stereographic projection from a celestial sphere onto a plane~\cite{Grenzebach:2014fha}. Note that the ``local shadow'' proposed in Ref.~\cite{Abdolrahimi:2015rua} is different from ours. Our notations here are similar to ones in Ref.~\cite{Luminet:1979nyg}.

A unit celestial sphere for the observer ${\rm O}'$ is drawn in Fig.~\ref{fg:angles}(b). In this figure, the $+z$ direction is in the direction of ${\rm O}$. We define the celestial coordinates $\left( \alpha , \beta \right)$ as in Fig.~\ref{fg:angles}(b), by which the incident angles of photon into the observer are specified. Note that angle $\alpha$ in Fig.~\ref{fg:angles}(b) corresponds to $\alpha = \angle {\rm X'OW}$ in Fig.~\ref{fg:angles}(a), which is determined by the positions of the observer and emitting particle, {\it i.e.}, by $i$ and $\psi$ [see Eqs.~\eqref{eq:alpha_gamma} and \eqref{eq:gamma_psi}].

Applying the law of sines in the spherical trigonometry to the spherical triangles ${\rm EXX'}$ and ${\rm EYY'}$, we have
\eq{
	\frac{ \sin (\pi/2) }{ \sin( \pi/2-\psi ) }
	=
	\frac{ \sin \xi }{ \sin( \pi/2-\alpha ) },
\;\;\;
	\frac{ \sin \xi }{ \sin( \pi/2-i ) }
	=
	\frac{ \sin (\pi/2) }{ \sin \gamma }.
\label{eq:law_of_sines}
}
Here, $\xi = \angle {\rm YEY'}$, and $\gamma = \angle {\rm EOO'}$. For simplicity, let us call $\gamma$ the deflection angle, while the standard deflection angle in gravitational lensing phenomena corresponds to $\pi -2\gamma$. Eliminating $\xi$ from two relations in Eq.~\eqref{eq:law_of_sines}, one obtains
\eq{
	\cos \alpha
	=
	\frac{\cos \psi \cos i}{ 
	\sin \gamma }.
\label{eq:alpha_gamma}
}

Another relation among angles we need is
\eq{
	\cos \gamma = \sin i \cos \psi,
\label{eq:gamma_psi}
}
which is obtained by the following elementary geometric consideration. First, if one draws a perpendicular from ${\rm Y'}$ to ${\rm OE}$ and calls its foot ${\rm H}$, ${\rm OH} = \cos \gamma$. Next, if one draws a perpendicular from ${\rm Y'}$ to ${\rm OY}$ and calls its foot ${\rm K}$, $\triangle {\rm HKY'}$ is a right triangle with $\angle {\rm HKY' }= \pi /2$. Using this fact, one can represents ${\rm OH}$ in another way, as ${\rm OH} = \sin i \cos \psi$. Thus, we obtain relation \eqref{eq:gamma_psi}.

Eliminating $\psi$ from Eqs.~\eqref{eq:alpha_gamma} and \eqref{eq:gamma_psi}, one obtains
\eq{
	\cos \alpha = \cot i \cot \gamma.
\label{eq:alpha_i_gamma}
}

To calculate $\gamma$, we identify the equatorial plane $\theta= \pi /2$ in the Schwarzschild coordinates of Eq.~\eqref{eq:line_element} with plane ${\rm OX'Y'}$ in Fig.~\ref{fg:angles}(a) and further identify $\phi$ in Eq.~\eqref{eq:line_element} with the angle of photon's position measured from ${\rm OX'}$. After this identification, we set
\eq{
	r_1=r_e, \;\; 
	\phi(r_1)= \frac{\pi}{2} -\gamma, \;\;
	r_2 = r_o, \;\;
	\phi(r_2) = \frac{\pi}{2}
}
in Eq.~\eqref{eq:phi_int}. Then,  we obtain
\eq{
	\gamma = \int_{r_e}^{r_o} \frac{b}{ r \sqrt{ r^2- b^2 f(r) }} dr.
\label{eq:gamma_int}
}

Thus, using Eqs.~\eqref{eq:alpha_i_gamma} and \eqref{eq:gamma_int}, we obtain the one of the incident angle, $\alpha$, as a function of $i, M, r_o, r_e$, and $b$.

Now, let us get down to how to calculate the rest incident angle, $\beta$. The tangent of $\beta$ is given by
{\eq{
	\tan \beta
	=
	\left. \frac{ p_{(\phi)} }{ p_{(r)} } \right|_{(r, \theta)=(r_o, \pi /2)},
\label{eq:tanbeta}
}
where $p_{(a)} := e_{(a)}^\mu p_\mu \;  (a=t,r,\theta,\phi)$ is the tetrad component of the photon's momentum. The tetrad basis $e_{(a)}$ is given by
\eq{
	e_{(t)} = \frac{1}{\sqrt{f(r)}} \pd_t,
\;\;\;
	e_{(r)} = \sqrt{f(r)} \pd_r,
\;\;\;
	e_{(\theta)} = \frac{1}{r} \pd_\theta,
\;\;\;
	e_{(\phi)} = \frac{\csc \theta}{r} \pd_\phi.
\label{eq:tetrad}
}
Note that $w := e_{(t)} |_{r=r_o}$ is also the 4-velocity of the observer. Using Eqs.~ \eqref{eq:tphidot}, \eqref{eq:drdlambda'}, \eqref{eq:tanbeta}, and \eqref{eq:tetrad} and the fact of $p_r = \frac{\pd \cal L}{ \pd \dot{r} }  = f(r)^{-1} \dot{r}$, we obtain
\eq{
	\tan \beta
	= \sqrt{ \frac{ b^2 f(r_o) }{ r_o^2-b^2 f(r_o) } }.
	\label{eq0240}
}
Thus, we obtain $\beta$ as a function of $M, r_o$, and $b$.

As in Fig.~\ref{fg:angles}(b), point ${\rm Q}$ on the unit celestial sphere,  specified by $(\alpha,\beta)$, is projected onto point ${\rm P}$ on the plane (a photographic plate) specified by $(x,y)$ in a stereographic way~\cite{Grenzebach:2014fha}. The relations between these coordinates are
\begin{eqnarray}
	x = -2 \sin \alpha \tan \frac{\beta}{2},
\;\;\;
	y = -2 \cos \alpha \tan \frac{\beta}{2}.
	\label{eq0320}
\end{eqnarray}
A circle with radius $2$ in the $x$-$y$ plane corresponds to the celestial equator.

In summary, we can obtain the two-dimensional image of the accretion disk (therefore, also the black hole shadow and apparent shape of the black hole) using Eqs.~\eqref{eq:alpha_i_gamma}, \eqref{eq0240}, and \eqref{eq0320} for a fixed values of $(M,r_o,i)$ by changing $(r_e,b)$ within the whole range allowed. 

%-------------------------------------------%
\subsection{Map from the parameter space to apparent-shape library}
\label{sec:mappin}
%-------------------------------------------%

For our purpose to examine whether one can determine the parameters of (black hole) + (accretion disk) + (observer) system by observation, it is convenient to define a parameter space ${\cal P}$, an apparent-shape library ${\cal I}$ (letter ${\cal I}$ stands for image), and a map $\varPsi : {\cal P} \to {\cal I}$~\cite{Hioki:2009na}.  

Parameter space ${\cal P}$ in our problem is defined by
\begin{eqnarray}
	\mathcal{P}
	:=
	\{
		\left( M, r_{o}, i \right)
			\mid 
		M>0, r_{o}  \geq 20M, 0 \leq i < \pi /2
	\}  \subset {\mathbb R}^3.
\end{eqnarray}
For later convenience, we also define an equivalence relation $\sim$ in ${\cal P}$ by
\begin{eqnarray}
	\left( M, r_{o}, i \right) \sim \left( M', r_{o}', i' \right)
	\stackrel{\rm def.}{\iff}
	\begin{cases}
		r_{o}/M = r_{o}'/M' \\
		i = i' 
	\end{cases} .
\end{eqnarray}

Map $\varPsi$ is defined to map $(M,r_o,i) \in {\cal P}$ to an apparent shape in the way described in the final paragraph of Sec.~\ref{sec:image}. Then, apparent-shape library ${\cal I}$ is defined as the image of $\varPsi$ as ${\cal I} := \varPsi ({\cal P})$, namely, a collection of all apparent shapes that the present (black hole) + (accretion disk) + (observer) system generates in the prescribed way.

While map $\varPsi$ is surjective by definition, it is not necessarily injective. If $\varPsi$ is not injective, an element in ${\cal I}$ can correspond to two or more distinct elements in ${\cal P}$, which means that one cannot determine the parameters from an apparent shape. One the other hand, if $\varPsi$ is injective, $\varPsi$ is bijective or invertible so that one can always uniquely specify black hole parameters $(M,r_o,i) \in {\cal P}$ corresponding to an apparent shape in ${\cal I}$, that is identical to (or approximating enough in reality) an actual observed apparent shape.

%-------------------------------------------%
\section{Size and shape of shadow}
\label{sec:shape}
%-------------------------------------------%

%-------------------------------------------%
\subsection{Deflection angle and images}
\label{sec:deflection}
%-------------------------------------------%

What is left before drawing the two-dimensional image is to calculate the integral in $\gamma$ [see Eq.~\eqref{eq:gamma_int}]. Although the integral cannot be written down in terms of elementary functions unfortunately at least in the case that $r_o$ is infinite (see Ref.~\cite{Chandrasekhar:1985kt}, p.132 and 134), it is known that the integral can be written down in terms of the incomplete elliptic integral of the first kind,
\eq{
	{\sf F}(\varPhi, K) := \int_0^\varPhi \frac{ d\vartheta }{ \sqrt{1-K^2 \sin^2 \vartheta} }.
\label{eq:gamma_first}
}
As will be seen soon, this is the case also when $r_o$ is finite.

It is convenient to introduce a parameter that labels the geodesics instead of impact parameter $b$. Such a parameter for the null geodesic of the first kind is the perihelion of orbit $P$, related to the impact parameter $b$ by
\begin{align}
	b^2 = \frac{P^3}{P-2M},
\;\;\;
	P>3M \;\;\; (\Leftrightarrow b>b_c).
\label{eq0300}
\end{align}
With this parameter, deflection angle $\gamma$ in Eq.~\eqref{eq:gamma_int} for finite $r_o$ is calculated to yield
\begin{eqnarray}
	\gamma = 2 \sqrt{\frac{P}{Q}}\Big[ {\sf F}\left( \zeta(r_{e}) , k  \right) - {\sf F}\left( \zeta(r_{o}) , k  \right) \Big].
\label{eq0250}
\end{eqnarray}
Here, all quantities in Eq.~\eqref{eq0250} are written in terms of $(M, r_o, r_e, P)$ through
\begin{gather}
	Q^2 := \left( P -2M  \right) \left( P+6M  \right),
\;\;\;
	k^2 := \frac{Q-P+6M}{2Q}, \\
	\sin ^2 \zeta(r) := \frac{Q-P+2M+4MP/r}{Q-P+6M}.
\end{gather}

For the null geodesic with an imaginary eccentricity, a convenient parameter that labels the geodesics instead of impact parameter is $\mu$ defined by
\eq{
	b^2 = \frac{M^2}{\mu (4 \mu -1)^2},
\;\;\;
	\mu > \frac13 \;\;\; (\Leftrightarrow 0<b<b_c).
}
With this parameter, deflection angle $\gamma$ in Eq.~\eqref{eq:gamma_int} for finite $r_o$ is calculated to yield
\begin{eqnarray}
	\gamma
	=
	\frac{1}{\sqrt{\varDelta}}
	\Big[
			{\sf F} \left( \sigma(r_{o}) , \kappa  \right) - {\sf F} \left( \sigma(r_{e}) , \kappa  \right)
	\Big].
	\label{eq:gamma_ie}
\end{eqnarray}
All quantities in Eq.~\eqref{eq:gamma_ie} are written in terms of $(M,r_o,r_e,\mu)$ through
\begin{gather}
	e^2 := \frac{3\mu -1}{\mu}, \;\;\;
	\varDelta ^2 := 48 \mu ^2 -16 \mu +1, \;\;\;
	\kappa^2 := \frac{\varDelta +6 \mu -1}{2 \varDelta}, \\
	\tan \frac{\xi(r) }{2} := \frac{1}{e} \left( \frac{M}{ \mu r}-1 \right), \;\;\;
	\sin ^2 \sigma(r)  := \frac{\varDelta -2\mu e \sin \xi(r) - (6\mu -1) \cos \xi(r)  }{\varDelta +6 \mu -1}.
\label{eq0360}
\end{gather}

Substituting Eqs.~\eqref{eq0250} and \eqref{eq:gamma_ie} into Eq.~\eqref{eq:alpha_i_gamma}, and using Eqs.~\eqref{eq0240} and \eqref{eq0320}, one can draw a two-dimensional image for a given set of values $(M,r_o,i)$ by changing $r_e \in [6M,20M]$, $P$, and $\mu$.

%-------------------------------------------%
\begin{figure}[htp]
		\begin{tabular}{ cccc }
			\includegraphics[width=3.8cm]{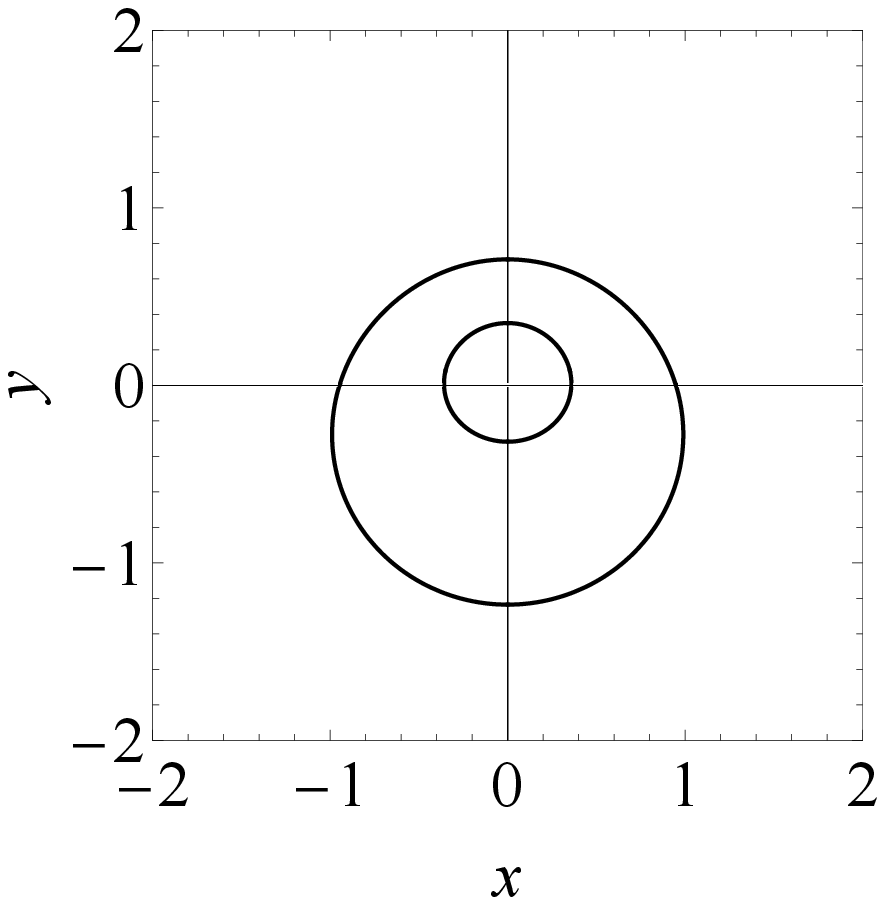} &
			\includegraphics[width=3.8cm]{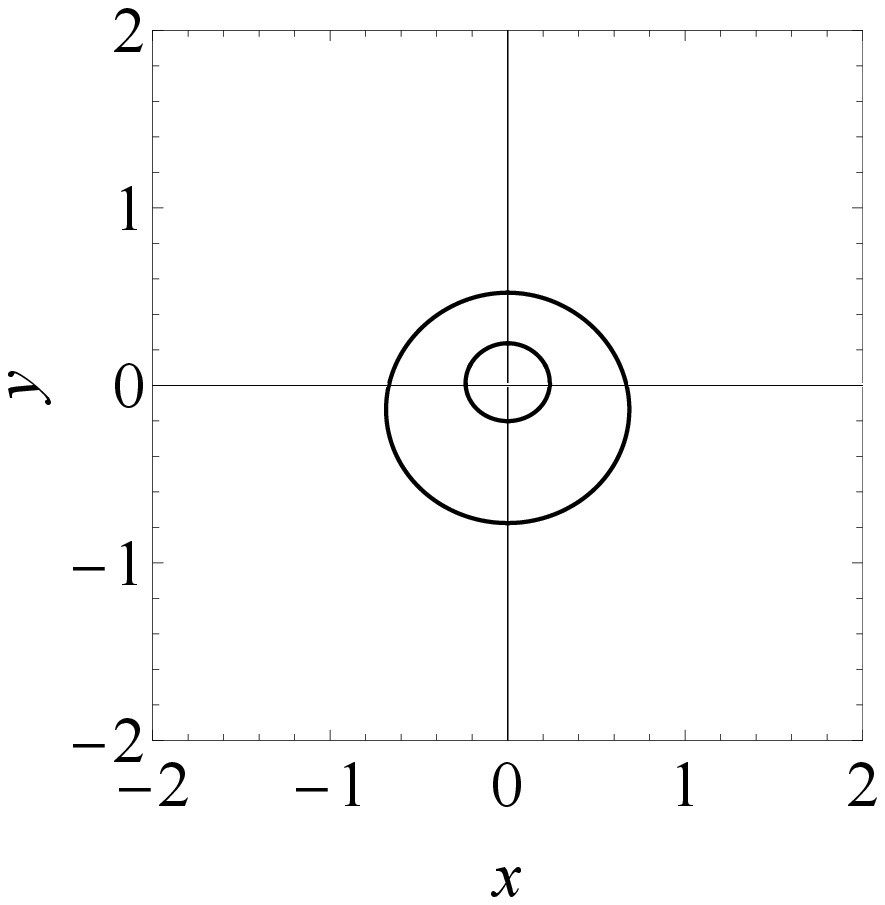} &
			\includegraphics[width=3.8cm]{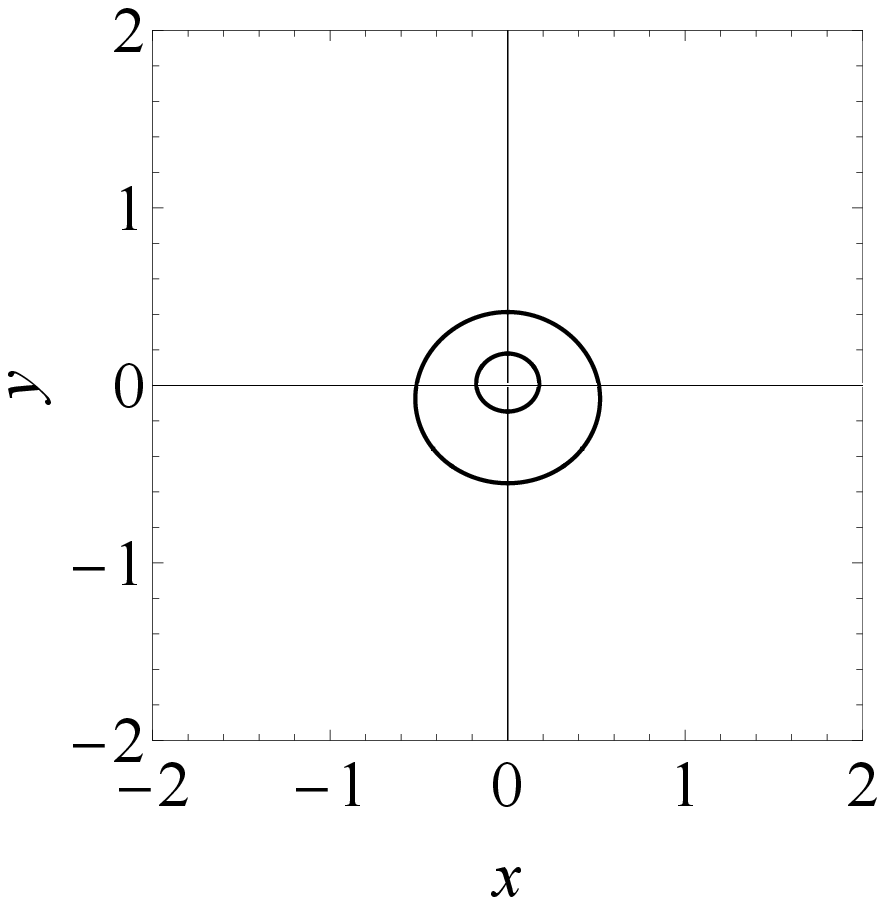} &
			\includegraphics[width=3.8cm]{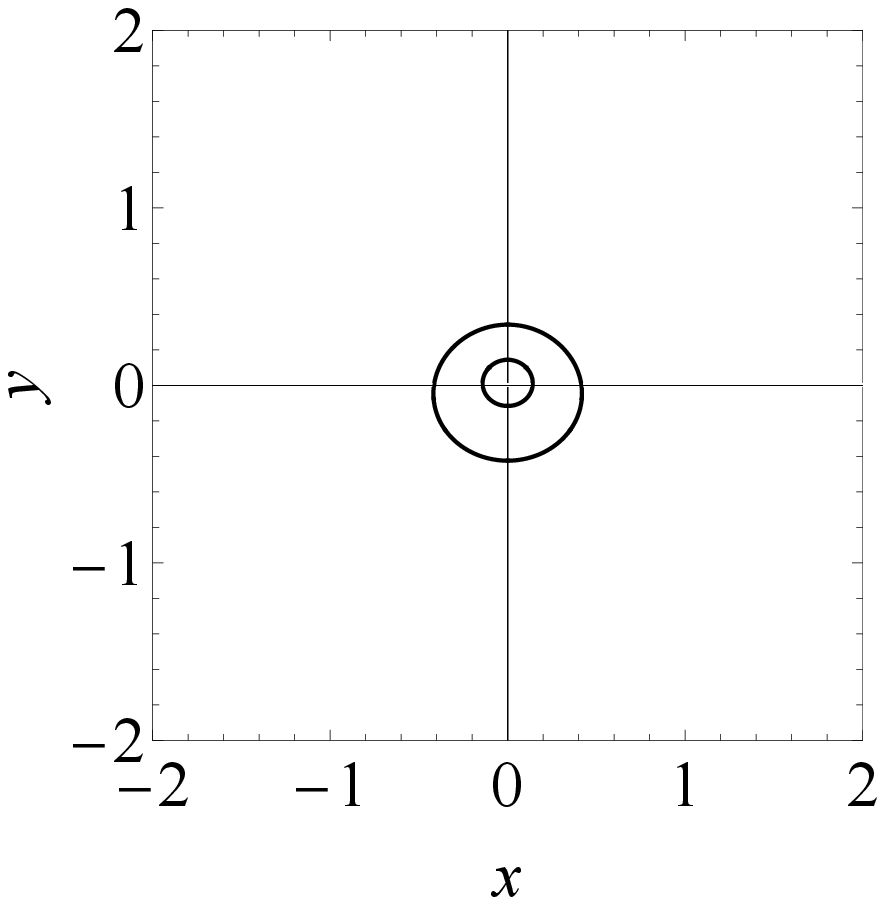} \\
			(a) $r_{o}/M =20 $, $i=30^\circ$ &
			(b) $r_{o}/M =30 $, $i=30^\circ$ &
			(c) $r_{o}/M =40 $, $i=30^\circ$ &
			(d) $r_{o}/M =50 $, $i=30^\circ$ \\
			\includegraphics[width=3.8cm]{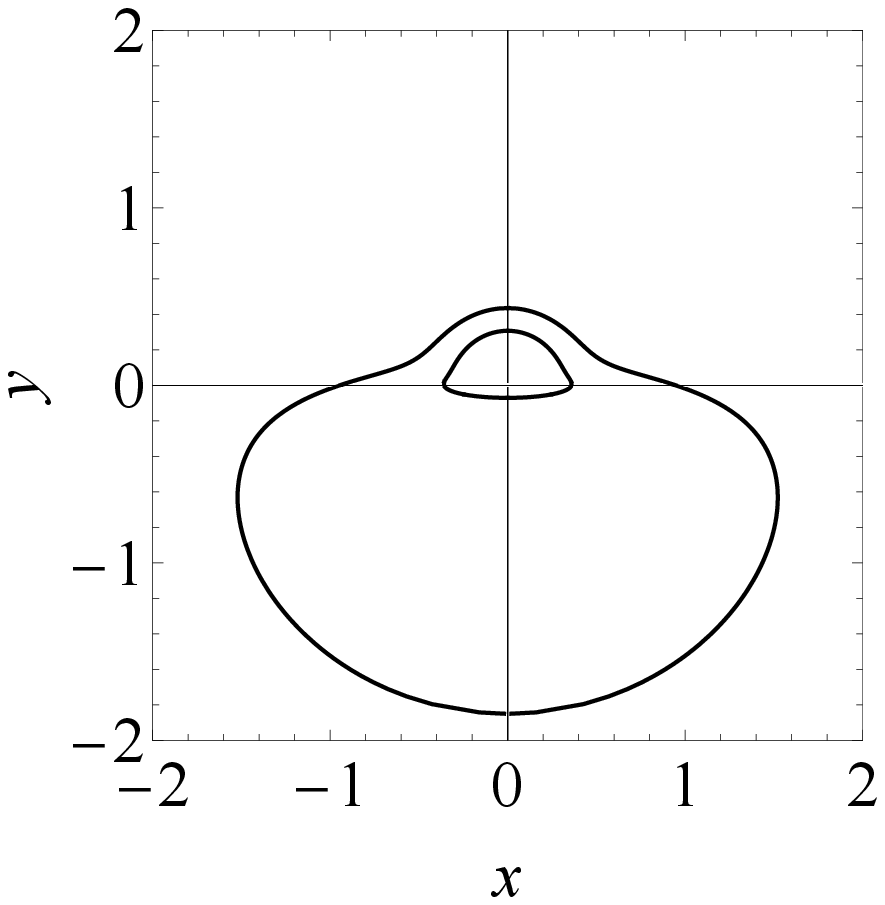} &
			\includegraphics[width=3.8cm]{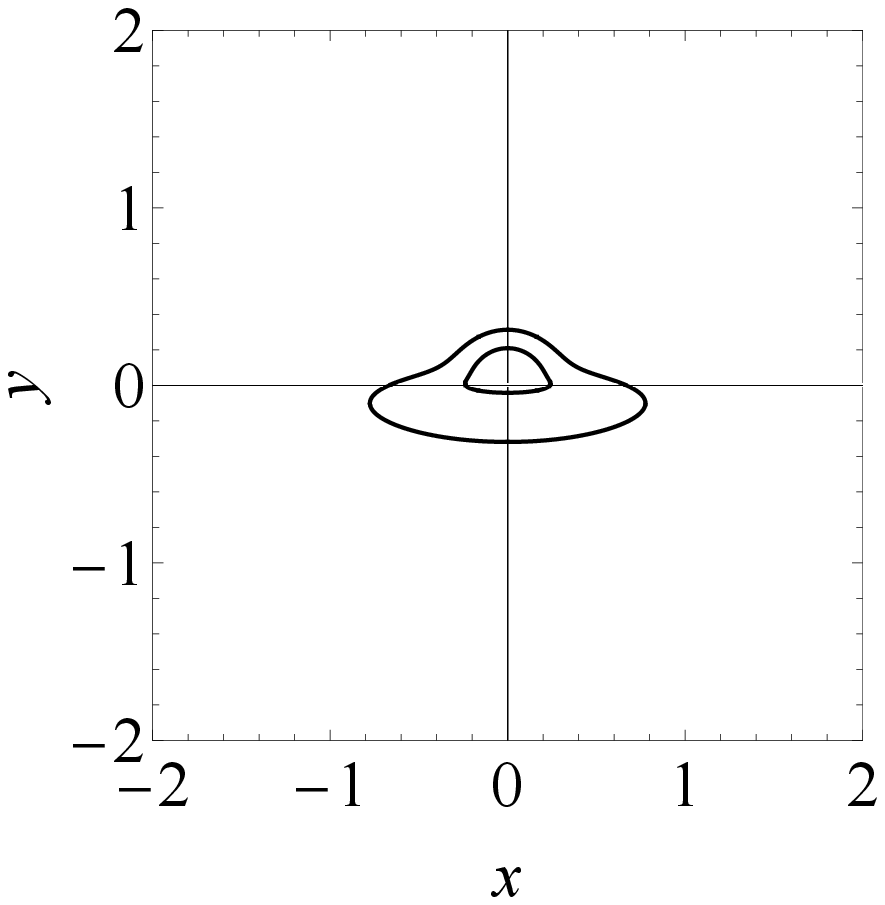} &
			\includegraphics[width=3.8cm]{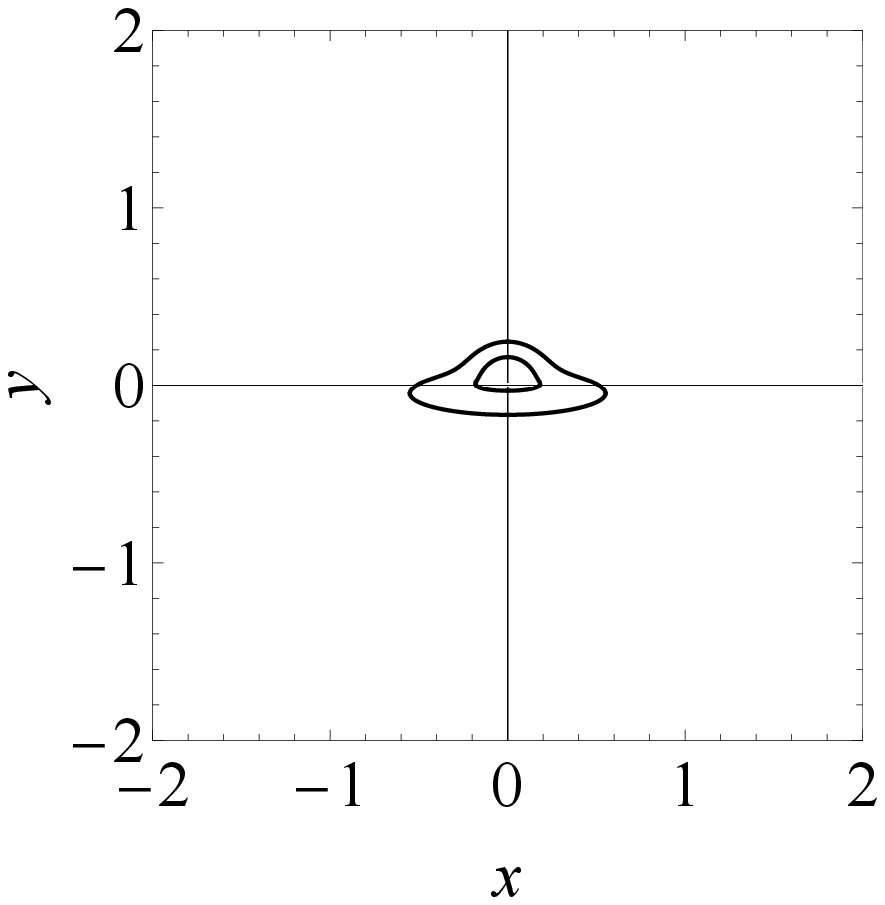} &
			\includegraphics[width=3.8cm]{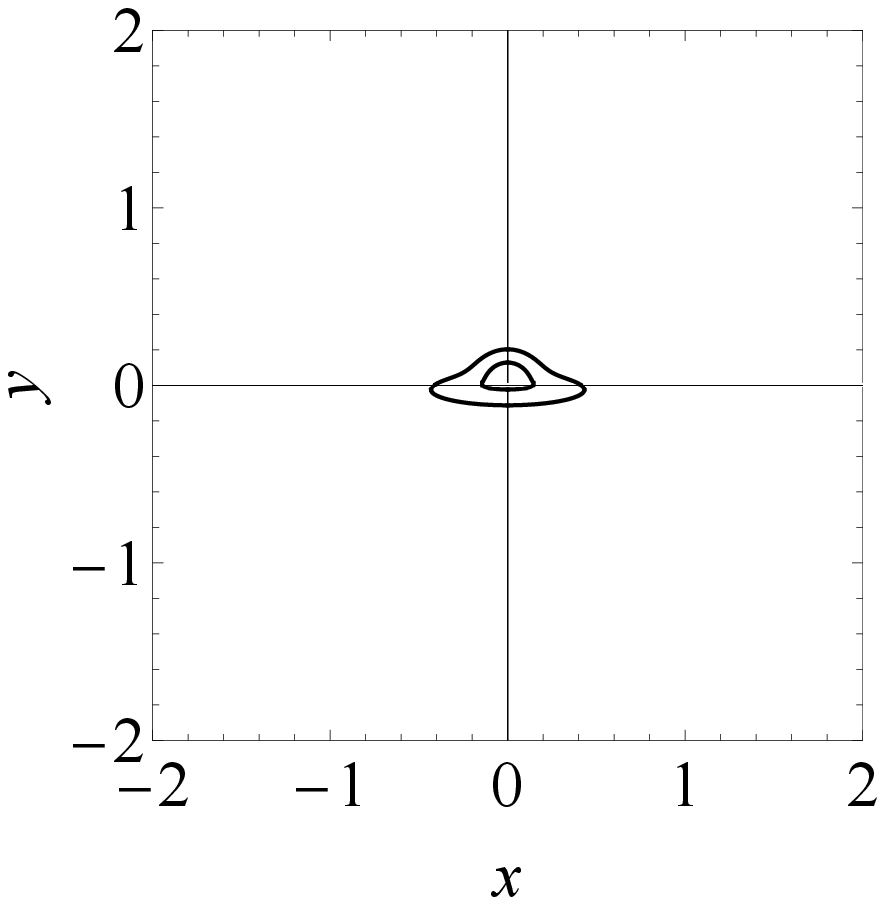} \\
			(e) $r_{o}/M =20 $, $i=80^\circ$ &
			(f) $r_{o}/M =30 $, $i=80^\circ$ &
			(g) $r_{o}/M =40 $, $i=80^\circ$ &
			(h) $r_{o}/M =50 $, $i=80^\circ$
		\end{tabular}
\caption{\footnotesize{Shadows of Schwarzschild black holes surrounded by thin accretion disks.} }				
		\label{fig01}
\end{figure}
%-------------------------------------------%

In Fig.~\ref{fig01}, we present the images for two values of inclination angle $i=30^\circ$ and $80^\circ$ and four values of dimensionless distance $r_o/M = 20, 30, 40$, and $50$. In each figure, we only plot the apparent shape of the accretion disk, which is generated by the null rays emitted from the inner and outer edges of accretion disk. The inner boundary of the apparent shape of the accretion disk is nothing but the apparent shape of the black hole by definition.

When the inclination angle is small, we see that the disk seems like a slightly deformed annulus, which is the well-known result~\cite{frolov2011introduction}, and the apparent size of such an annulus becomes small as $r_o$ increases as expected.

Even when the inclination is large, the ``opposite'' side of the accretion disk far from the observer is always visible due to the strong deflection of light rays by the black hole, as is well known. An interesting feature of the apparent shape appears when the inclination angle is large and $r_o /M$ is close to $20$. Namely, in such cases, the annulus is highly deformed.
The fact that the apparent shapes seen by observers at infinity and at finite distance are not similar (namely, not having an identical shape) is important for the determination of physical quantities.

The reason why we have presented the images in Fig.~\eqref{fig01} for several values of $r_o/M$ is that the size and shape depend on $r_o/M$, of which the inverse is sometimes called the {\it angular gravitational radius}, rather than $r_o$ and $M$ themselves. This can be proved rigorously. Namely, the position of image $(x, y)$ is the function of $(M,r_e,r_o,i,b)$ (or $P$ and $\mu$ instead of $b$) through Eqs.~\eqref{eq:alpha_i_gamma}, \eqref{eq0240}, \eqref{eq0320}, \eqref{eq0250}, \eqref{eq:gamma_ie}, and so on. However, if we normalize every dimensionful quantity by $M$ as
\eq{
	r_{e\ast} := \frac{r_{e}}{M},
\;\;\;
	r_{o\ast} := \frac{r_{o}}{M}, 
\;\;\;
	b_\ast := \frac{b}{M}, 
\;\;\;
	P_\ast
	:=
	\frac{P}{M} ,
}
the position of image $(x,y)$ becomes the function of dimensionless quantities $(i, r_{e\ast}, r_{o\ast}, b_\ast)$ (or $P_\ast$ and $\mu$ instead of $b_\ast$) and does not depend on $M$ explicitly. In other words, when $(r_o/M, i) = (r_o/M', i')$ holds, the image of system with $(M, r_o, i)$ and that of the system with $(M', r_o', i')$ are completely {\it congruent}. 

From the above fact, we can say that all elements in an equivalence class $[(M, r_o, i)]$ of quotient space ${\cal P}/\sim$ are mapped to an image or shape in ${\cal I}$ by a map $\varPsi$. This does not mean, however, that a map $\varPsi _q$ is injective, where $\varPsi _q$ is defined as $\varPsi _q: {\cal P}/\sim \ni [(M, r_o, i)] \mapsto \varPsi((M, r_o, i)) \in {\cal I}$. In the next subsection, we will show that $\varPsi _q: {\cal P}/\sim \; \to {\cal I}$ is indeed injective, which means that one can determine value of $(r_o/M,i)$ by capturing the image.

%-------------------------------------------%
\subsection{Size and shape determines $(r_o/M, i)$}
\label{sec:determinestepone}
%-------------------------------------------%

Since the ISCO has a universal meaning as we saw in Sec.~\ref{sec:timelike}, we focus on the apparent shape of the black hole corresponding to the ISCO.
Then, we will define two observables characterizing the apparent shape of the black hole~\cite{Hioki:2009na}.

We approximate the apparent shape of the black hole by a circle passing through the three points located at the top position, the leftmost end, and the rightmost end of the shadow as in the three green points in Fig.~\ref{fig02}. We denote the radius of this circle by $R$.

Next, let us consider the dent in the bottom part of the shadow. The size of the dent is denoted by $D$, which is the length between the bottom positions of circle and the shadow as in Fig.~\ref{fig02}. Then, we define the distortion parameter of the shadow by $\delta := D/R$.

%-------------------------------------------%
\begin{figure}[htp]
		\begin{tabular}{ c }
			\includegraphics[height=5cm]{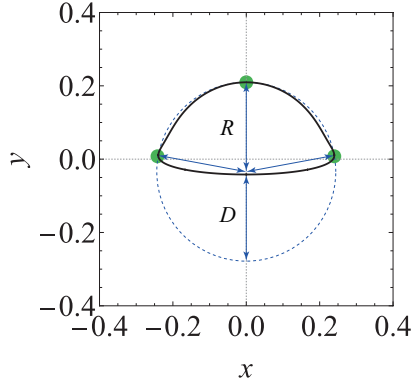}
		\end{tabular}
	\caption{\footnotesize{The observables for the apparent shape of a Schwarzschild black hole surrounded by a thin disk are the radius $R$, being the radius of an approximating circle, and the distortion parameter $\delta := D/ R$, where $D$ is the difference between the down end points of the circle and of the shadow.}}
	\label{fig02}
\end{figure}
%-------------------------------------------%

We present the contour plots of $R$ and $\delta$ in Fig.~\ref{fig03}(a) and \ref{fig03}(b), respectively. The contour plots of $R$ and $\delta$ are overlayed in Fig.~\ref{fig04}. From Fig.~\ref{fig03}, we can see that $R$ decreases as $r_o/M$ increases, and $\delta$ increases as $i$ increases, which we have already seen in Fig.~\ref{fig01}. Note that the apparent shape is so distorted that $\delta$ exceeds $1$ for $i \gtrsim 80^\circ $. 

Here, the following two facts are important for our purpose:
\begin{itemize} 
\item
From Fig.~\ref{fig03}(b), even when $i = {\rm const.}$, $\delta  \neq {\rm const.}$
\item
From Fig.~\ref{fig04}, there is a one-to-one correspondence between ($r_{o}/M, i$) and ($R, \delta$).
\end{itemize}

The former means that two apparent shapes with distinct values of $r_o/M$ are {\it not similar} even if the inclination angle $i$ is common. The latter means that map $\varPsi _q: \mathcal{P}/\sim \; \to \mathcal{I}$ is invertible (bijective). Therefore, if one measures $R$ and $\delta$ by observations, the values of $r_o/M$ and $i$ are determined by using Fig.~\ref{fig04}.

%-------------------------------------------%
\begin{figure}[tb]
		\begin{tabular}{ cc }
			\includegraphics[height=5cm]{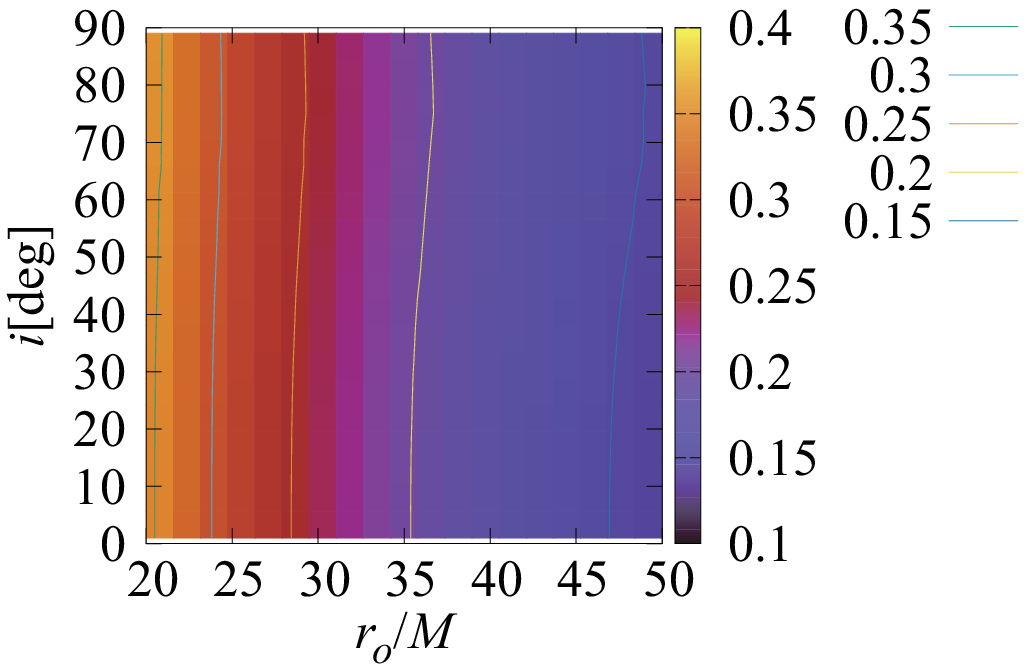} &
			\includegraphics[height=5cm]{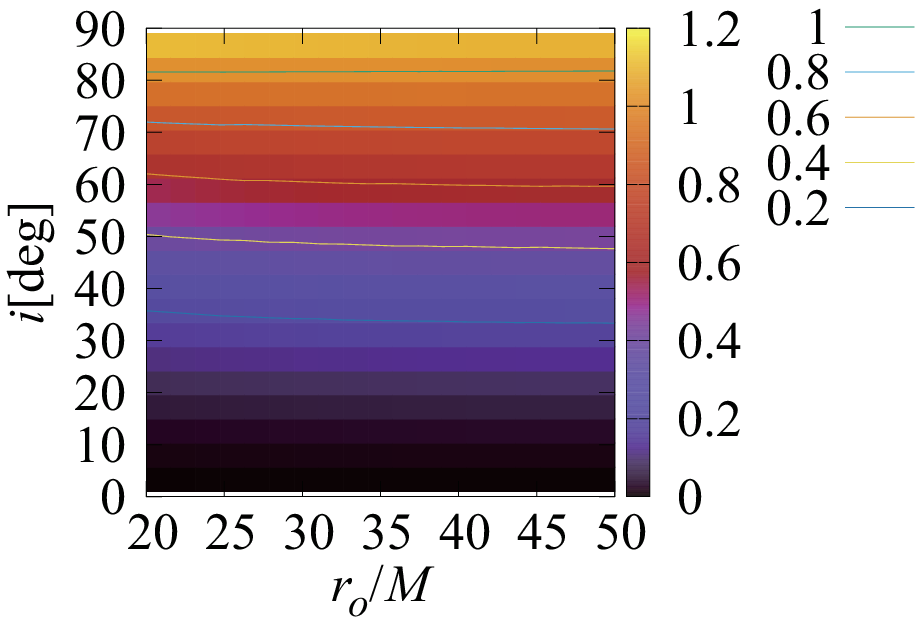} \\
			(a) & (b) \\
		\end{tabular}
	\caption{\footnotesize{(a) A contour map of the radii of shadow $R$ of for the Schwarzschild black hole surrounded by the thin accretion disk. (b) A contour map of the distortion parameter $\delta$.}}
	\label{fig03}
\end{figure}
%-------------------------------------------%

%-------------------------------------------%
\begin{figure}[tb]
		\begin{tabular}{ c }
			\includegraphics[height=5cm]{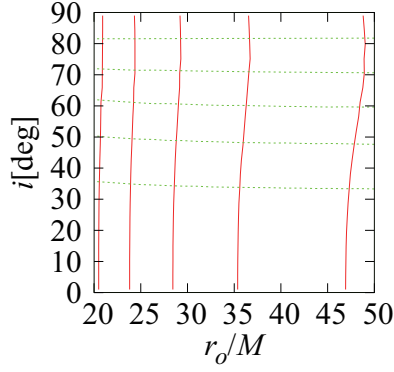}
		\end{tabular}
	\caption{\footnotesize{The contours of the radius of shadow $R$ (the red solid curves) and the distortion parameter $\delta$ (the green dashed curves). The contours are those for $R = 0.15, \cdots, 0.35$, with the contour interval being 0.05, and $\delta = 0.2, \cdots,1$, with the contour intervals being 0.2.}}
	\label{fig04}
\end{figure}
%-------------------------------------------%

%-------------------------------------------%
\section{Flux of the accretion disk}
\label{sec:flux}
%-------------------------------------------%

We have seen that information read from the apparent shape of the shadow is insufficient to determine the parameters completely. Therefore, in this section, we utilize another observationally measurable quantity, {\it i.e.}, the energy flux emitted by the accretion disk, in order to make the parameter determination complete.

%-------------------------------------------%
\subsection{Radial dependence of bolometric flux}
\label{sec:depflux}
%-------------------------------------------%

Under quite general assumptions, Page and Thorne derived the radial dependence of energy flux radiated by a thin accretion disk around a stationary axially symmetric black hole in an explicit form~\cite{Page:1974he, Abramowicz:2011xu}. Restricting their general result to the present Schwarzschild case~\cite{Luminet:1979nyg}, the bolometric energy flux flowing out of the upper (or equally the lower) face of the accretion disk is given as a function of emitting point $r_{e\ast} = r_e/M$, by
 \begin{eqnarray}
	F_e = \frac{3\dot{M}}{8 \pi M^2} \frac{1}{ ( r_{e\ast} -3 ) r^{5/2}_{e\ast}} \left( \sqrt{r_{e\ast}} - \sqrt{6} + \frac{\sqrt{3}}{2} \ln \left[  \frac{ ( \sqrt{6} - \sqrt{3}  ) ( \sqrt{r_{e\ast}} + \sqrt{3} )  }{ ( \sqrt{6} + \sqrt{3}  ) ( \sqrt{r_{e\ast}} - \sqrt{3} ) }  \right] \right),
	\label{eq520}
\end{eqnarray}
where $\dot{M} \; (>0)$ is the radius-independent mass accretion rate.
Note that the flux $F_e$ of light emitted from the ISCO is zero, as can be seen from Eq.~\eqref{eq520}.
The bolometric flux $F_o$ observed by an observer away from the disk differs from the above intrinsic flux $F_e$ by the inverse fourth power of redshift factor $1+z$~\cite{Ellis},
\begin{eqnarray}
	F_{o}  = \frac{F_e}{(1 +z)^4}.
\label{eq510}
\end{eqnarray}

In the present case, the redshift and/or blueshift consists of the Doppler effect due to the motion of emitter and the gravitational redshift, which is given by
\eq{
	1 + z 
	=
	\frac{- u^\mu p_\mu \mid _{r=r_e}}{- w^\mu p_\mu \mid _{r=r_o}},
\label{eq:redshift1}
}
where $u$ is the 4-velocity of emitting particle on the disk and $w$ is that of the observer at ${\rm O}'$ [see the sentence following Eq.~\eqref{eq:tetrad}].
Note that the following equations hold (see Sec.~\ref{sec:geodesics}):
\begin{gather}
	(p_t, p_r, p_{\theta}, p_{\phi})
	=
	( -E, f(r)^{-1}\sqrt{E^2-f(r)L^2/r^2}, 0 , L ),
\;\;\;
	p_{\bar{\phi}} = p_\phi \cos \xi,
\label{eq:pmu}
\\
	(u^t, u^r, u^{\bar{\theta}}, u^{\bar{\phi}})
	=
	(u^t, 0, 0, \varOmega u^{t}),
\;\;\;
	-1 = g_{\mu\nu}u^{\mu}u^{\nu}.
\label{eq:umu}
\end{gather}
Substituting Eqs.~\eqref{eq:pmu} and \eqref{eq:umu} into Eq.~\eqref{eq:redshift1} and using $\cos \xi = \sin i \sin \alpha$, which can be derived from Eqs.~\eqref{eq:law_of_sines} and \eqref{eq:gamma_psi}, we have the explicit form of redshift factor,
\eq{
	1 + z
	=
	\sqrt{\frac{1 - 2/r_{o\ast}}{1 -3/r_{e\ast}}}
	\left(
		1 - \frac{b_\ast}{ r_{e\ast}^{3/2} } \sin i \sin \alpha
	\right).
\label{eq:redshift2}
}

%-------------------------------------------%
\subsection{Map from the parameter space to image library}
\label{sec:mappin2}
%-------------------------------------------%

Now, we would like to upgrade the map from the parameter space to the apparent-shape library, $\varPsi : {\cal P} \to {\cal I}$, to another map $\hat{\varPsi} : \hat{\cal P} \to \hat{\cal I}$, in which not only the information about apparent shape but also that about the flux is involved.

First, we upgrade parameter space ${\cal P}$ to $\hat{\cal P}$ by
\begin{eqnarray}
	\hat{\cal P}
	:=
	{\cal P}
	\times
	\{
		\dot{M}
			\mid 
		\dot{M} > 0
	\}   \subset {\mathbb R}^4 .
\end{eqnarray}

Next, we consider a set of all possible two-dimensional images $\bar{\cal I}$, in which not only the information about the apparent shape of the black hole but also the information about the spatial distribution of bolometric flux on the accretion disk is contained. This set $\bar{\cal I}$ is regarded as the codomain of new map $\hat{\varPsi}$. Namely, map $\hat{\varPsi}$ is defined to map $(M, r_o, i, \dot{M}) \in \hat{\cal P}$ to an element (namely, a two-dimensional image) in $\bar{\cal I}$. 

Finally, $\hat{\cal I} \subset \bar{\cal I}$ is defined as the image of $\hat{\varPsi}$ as $\hat{\cal I} := \hat{\varPsi} ( \hat{\cal P} )$. Hereafter, we call $\hat{\cal I}$ the image library, which is a collection of all two-dimensional images that the present (black hole) + (accretion disk) + (observer) system generates in the way described in Secs.~\ref{sec:image} and \ref{sec:depflux}.

Remember that we change parameters $P_\ast$ and $\mu$ within the respective allowed range to plot the apparent shape of the accretion disk in Fig.~\ref{fig01}. Let $I \subset (3, r_{e}/M)$ and $J \subset ( 1/3 , \infty )$ be such a range of $P_{*}$ and $\mu$, respectively. Then, map $\hat{\varPsi}$ defined above can be written in a more specific way as
\begin{align}
	\hat{\varPsi}: & \left( M , r_{o}, i, \dot{M} \right) \in \hat{\cal P} \nonumber \\
	& \mapsto \{ \left( x , y, F_{o} \right) \mid P_\ast \in I, r_{e\ast} \in [6, 20] \} \sqcup \{ \left( x , y, F_{o} \right) \mid \mu \in J, r_{e\ast} \in [6, 20] \} \in \hat{\cal I},
	\label{eq530}
\end{align}
where $\sqcup$ stands for the disjoint union.

We also define an equivalence relation $\approx$ in $\hat{\cal P}$ by
\begin{eqnarray}
	\left( M, r_{o}, i, \dot{M} \right) \approx \left( M', r_{o}', i', \dot{M}' \right)
	\stackrel{\rm def.}{\iff}
	\begin{cases}
		r_{o}/M = r_{o}'/M' \\
		i = i' \\
		\dot{M}/M^2 = \dot{M}'/M'^2
	\end{cases} ,
\end{eqnarray}
which will be used in the next subsection.

%-------------------------------------------%
\subsection{Flux determines $(M, r_o, i)$}
\label{sec:determinesteptwo}
%-------------------------------------------%

We define a ``blueshifted flux'' $\mathcal{F}$ as an observable by	
\begin{eqnarray}
	\mathcal{F} := \{ \left( x , 0,  F_{o} \right) \mid \left( x , y , F_{o} \right) \in \hat{\cal I} , x < 0 , y = 0 , F_{o} > 0 \},
	\label{eq530}
\end{eqnarray}
which represents the $x$ dependence of observed flux $F_o$ along the $x$ axis with $x<0$.
See the yellow part in Fig.~\ref{fig07}. The reason why we focus on the $x<0$ region and call ${\cal F}$ the blueshifted flux is that the flux is blueshifted (enhanced) on this side due to the rotational motion of the accretion disk. Note that the origin and axes of $(x,y)$ coordinates can be identified by two observables $(R,\delta)$.

%-------------------------------------------%
\begin{figure}[htp]
		\begin{tabular}{ c }
			\includegraphics[height=5cm]{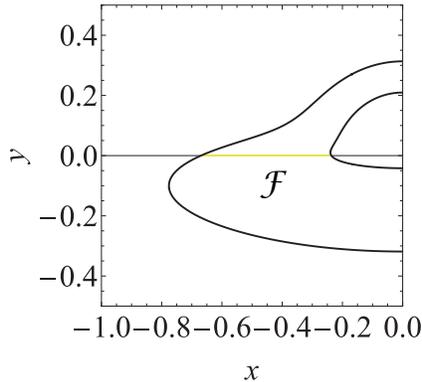}
		\end{tabular}
	\caption{\footnotesize{Blueshifted flux $\mathcal{F}$, representing the $x$ dependence of observed flux $F_o$ along the $x$ axis}}
	\label{fig07}
\end{figure}
%-------------------------------------------%

We also define a dimensionless blueshifted flux $\mathcal{F}_\ast$ by
\begin{eqnarray}
	\mathcal{F}_\ast := \left\{ \left( x , 0,  F_{o\ast} \right) \mid F_{o\ast} = \frac{ M^2 }{ \dot{M} } F_{o}, \left( x , 0 , F_{o} \right) \in \mathcal{F} \right\},
	\label{eq540}
\end{eqnarray}
which represents the $x$ dependence of dimensionless flux $F_{o\ast}$ along the $x$ axis with $x<0$. The dependence of $\mathcal{F}_\ast$ on $i$ and $r_o/M$ is shown in Fig.~\ref{fig05}. A three-dimensional plot is also given in Fig.~\ref{fig06}.

%-------------------------------------------%
\begin{figure}[htp]
		\begin{tabular}{ cccc }
			\includegraphics[width=3.4cm]{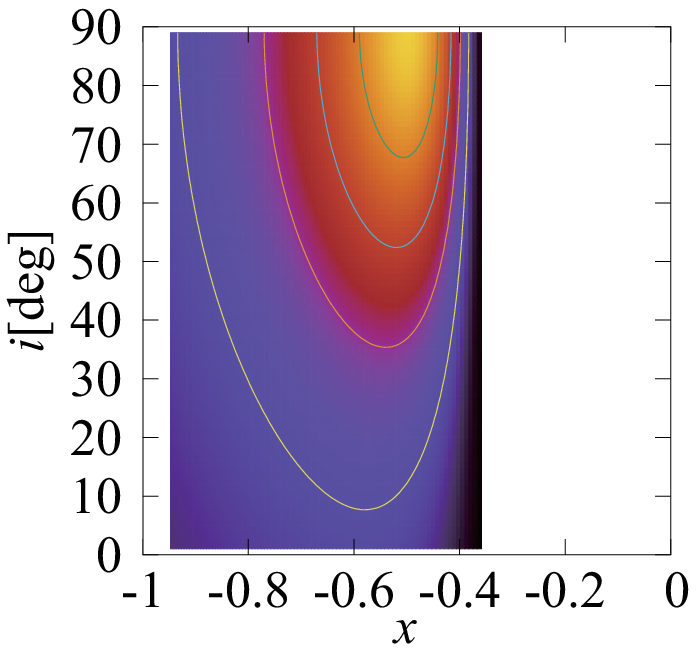} &
			\includegraphics[width=3.4cm]{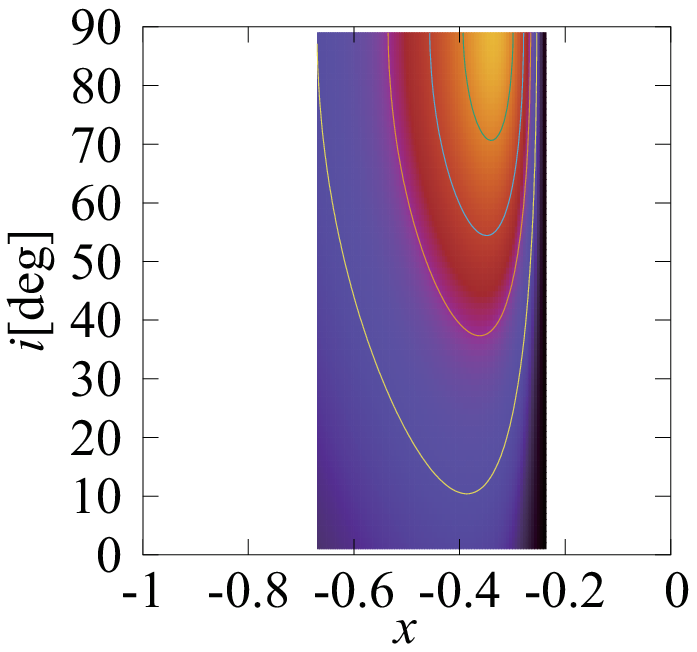} &
			\includegraphics[width=3.4cm]{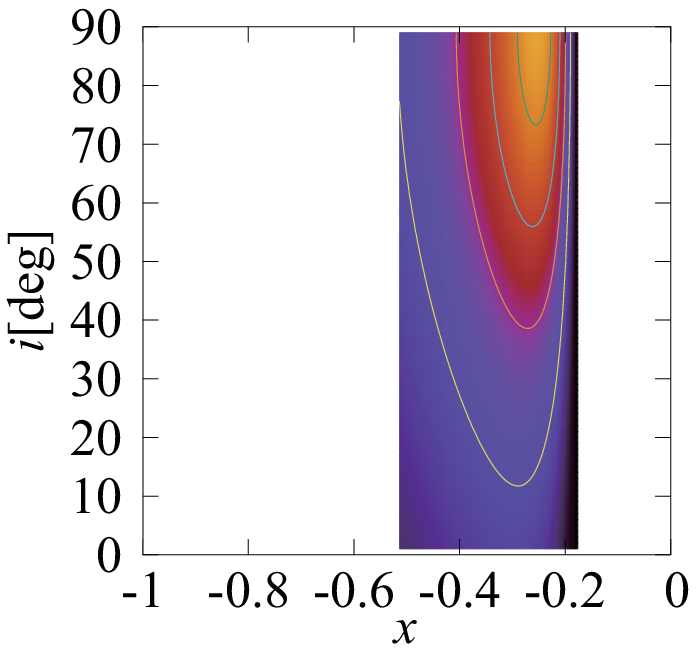} &
			\includegraphics[width=5.6cm]{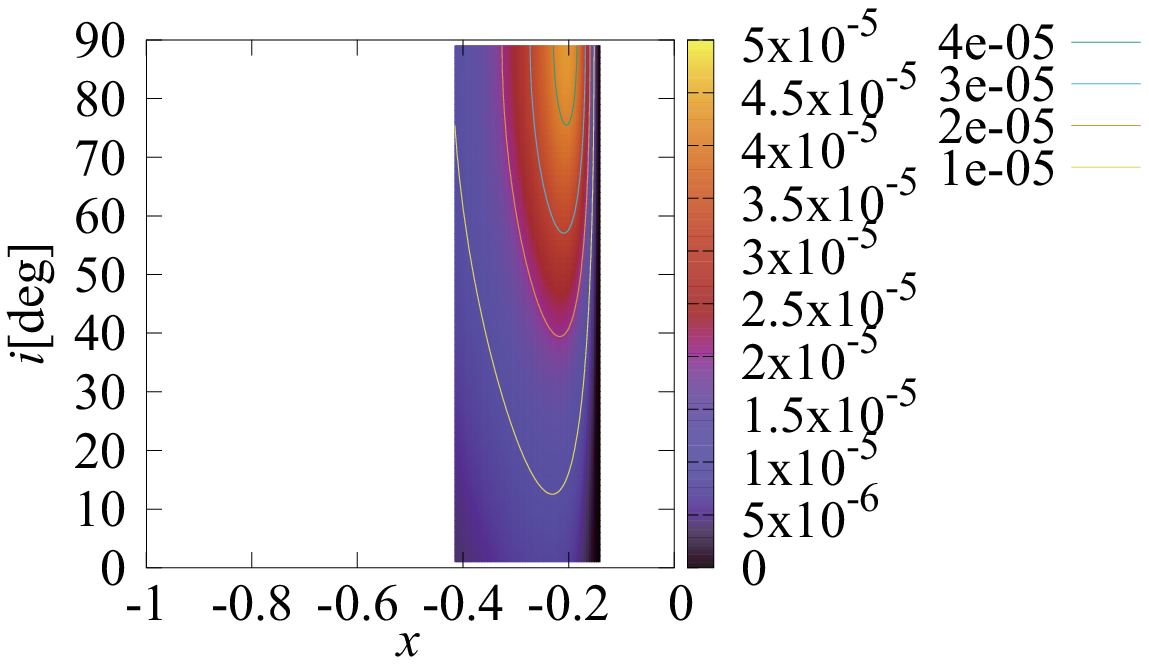} \\
			(a) $r_{o}/M =20 $ &
			(b) $r_{o}/M =30 $ &
			(c) $r_{o}/M =40 $ &
			(d) $r_{o}/M =50 $ \\
		\end{tabular}
	\caption{\footnotesize{Contour maps of dimensionless blueshifted flux $\mathcal{F}_\ast$.}}				
	\label{fig05}
\end{figure}
%-------------------------------------------%

%-------------------------------------------%
\begin{figure}[htp]
		\begin{tabular}{ c }
			\includegraphics[height=5.5cm]{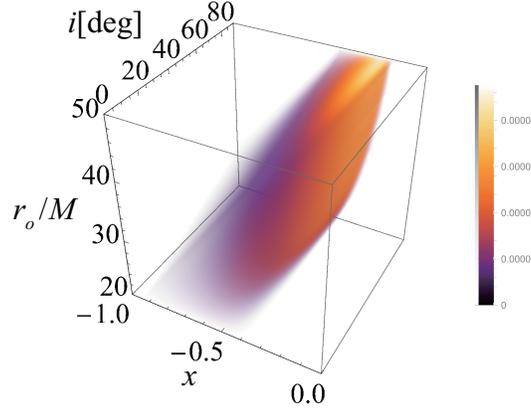}
		\end{tabular}
	\caption{\footnotesize{The dependence of dimensionless blueshifted flux $\mathcal{F}_\ast$ on $i$ and $r_o/M$.}}
	\label{fig06}
\end{figure}
%-------------------------------------------%

Now, let us suppose the following situations:
\begin{itemize}
\item[(i)]
From the observational measurement of $(R,\delta)$, we have already known $(r_o/M, i)$ of the system.

\item[(ii)]
We have observed the bolometric energy flux to know $F_o$ as a function of $(x,y)$ or as a function of $x$ on the $x$ axis with $x<0$ at least.
\end{itemize}

Under assumptions i and ii, parameters $(r_o/M , i, \dot{M}/M^2)$ can be determined only by capturing the two-dimensional image.
This is because the following relationship holds:
\eq{
	\frac{\dot{M}}{M^2}
	=
	\left.
		\frac{ F_o}{ F_{o\ast}}
	\right|_{x<0, y=0}.
\label{eq:mass_flux}
}
In other words, by observing the blueshifted flux ${\cal F}$ and using Figs.~\ref{fig05} and \ref{fig06} we can determine $\dot{M}/M^2$.
Note that the right-hand side of Eq.~\eqref{eq:mass_flux} apparently has the dependence on $x$ but it is constant and thus the right-hand side can be evaluated at {\it any} point on the $x$ axis, provided $x<0$ and $F_o>0$ at that point.  

We can say that all elements in an equivalence class $[(M, r_o, i, \dot{M})]$ of quotient space $\hat{\cal P}/\approx$ are mapped to a two-dimensional image in $\hat{\cal I}$ by a map $\hat{\varPsi}$.
From the fact that the map $\varPsi _q$ is bijective and Eq.~(\ref{eq:mass_flux}) holds, a map $\hat{\varPsi} _q$ is bijective, where $\hat{\varPsi} _q$ is defined as $\hat{\varPsi} _q: \hat{\cal P}/\approx \ni [(M, r_o, i, \dot{M})] \mapsto \hat{\varPsi}((M, r_o, i, \dot{M})) \in \hat{\cal I}$.
Namely, there exists a one-to-one correspondence between $\hat{\cal P}/\approx$ and $\hat{\cal I}$, and we can determine the parameters of the system, $(r_o/M , i, \dot{M}/M^2)$, only by capturing the two-dimensional image.

Furthermore, we suppose the following additional situation:
\begin{itemize}
\item[(iii)]
With any independent theoretical or observational approach, we know accretion rate $\dot{M}$ of the system.
\end{itemize}
Under three assumptions i, ii, and iii, we can estimate mass $M$.

In summary, we can say that it is possible to determine the parameters of the system, $(M, r_o , i)$, by capturing the two-dimensional image and from the information on the mass accretion rate.
Note that if $M$ is known (instead of $\dot{M}$), it is possible to determine $\dot{M}$.

%-------------------------------------------%
\subsection{Demonstration}
%-------------------------------------------%

Let us demonstrate how one obtains the value of $M$ from an observational data. Reviving $c$ and $G$ in Eq.~\eqref{eq540}, we have
\eq{
	F_{o \ast}
	=
	\frac{ G^2 M^2  }{ c^6 \dot{M} } F_{o}.
\label{eq:F_o*}
}
We assume that accretion rate $\dot{M}$ is $\epsilon$ (a dimensionless factor) times the critical or Eddington accretion rate $\dot{M}_E$, namely, $\dot{M} = \epsilon M_E := \epsilon L_E/c^2$, where $L_E$ is the Eddington luminosity, given by
\eq{
	L_E
	=
	\frac{ 4\pi cGm_p }{ \sigma_T } M,
\;\;\;
	\sigma_T
	=
	\frac{8\pi}{3}
	\left(
		\frac{ e^2 }{ 4\pi \varepsilon_0  m_e c^2}
	\right)^2.
}
Here, $\sigma_T $ is the cross section of Thomson scattering.  Then, from Eq.~\eqref{eq:F_o*} the mass is expressed as
\eq{
	M
	=
	\epsilon
	\cdot
	\frac{ 4\pi c^5 m_p }{  GM_\odot \sigma_T F_o }
	\cdot
	F_{o\ast} 
	M_\odot
	\simeq
	\epsilon \cdot \frac{ 5.76 \times 10^{25} \; {\rm erg/cm^2/s} }{ F_o } 
	\cdot
	F_{o\ast} 
	M_\odot.
\label{eq:M_estimate}
}
Suppose we have measured $(R, \delta)$ from the apparent shape of the black hole and gotten $(r_o/M, i) = (20, 30^\circ)$ by the method described in Sec.~\ref{sec:determinestepone}. Furthermore, we know that $\epsilon = 0.01$ by any independent approach and observationally know that flux $F_o$ on the $x$ axis takes a maximum as $(x,y,F_o) = (-5.494 \times 10^{-1}, 0, 1.536 \times 10^9{\rm erg/cm^2/s}) \in \mathcal{F}$. On the other hand, the theoretical value of $F_{o\ast}$ at the same position can be obtained by calculation as $(x,y,F_{o\ast}) = (-5.494 \times 10^{-1}, 0, 1.745 \times 10^{-5}) \in \mathcal{F}_{*}$ (see Fig.~\ref{fig05}). Substituting these values into the right-hand side of Eq.~\eqref{eq:M_estimate}, we obtain $ M \simeq 6.548 \times 10^9 M_\odot $.

%-------------------------------------------%
\section{Discussion}
\label{sec:discussion}
%-------------------------------------------%
Now, we would like to estimate how much the viewing angles differ between the existing and our new methods, regarding our important samples of M87 and ${\rm Sgr\; A^\ast}$ black holes. To do so, we need to carefully think about Bardeen coordinates $(b_x, b_y)$. The reason why the image of a black hole on Bardeen coordinates has a nonzero size, in spite of the black hole being assumed to be at spatial infinity, is that such an image is the so-called photon capture (see, e.g., Ref.~\cite{Hioki:2008zw}), which is an absorption cross section. 

The existing method based on Bardeen coordinates assumes that a light ray parallel to the line connecting the observer and the black hole reaches the observer, which is possible only when the distance is sufficiently large. Although the incident angle of light rays to the observer cannot be defined, the Bardeen coordinates can be converted to the celestial coordinates as $(b_x/r_o , b_y/r_o )$. So, the viewing angle is obtained on these celestial coordinates. 

On the other hand, in our new method, the incident angle is properly defined, and therefore the viewing angle of objects is calculated without any approximation. As the results, the viewing angle between the upper and lower points of the black hole shadow are calculated to be 48.8155 and 68.1276 microarcsecs for M87 and ${\rm Sgr\; A^\ast}$, respectively. Here, the inclination angle for M87 and ${\rm Sgr\; A^\ast}$ were assumed to be $20^\circ$ and $30^\circ$, respectively~\cite{EventHorizonTelescope:2019dse, EventHorizonTelescope:2022xnr}. 

From the above calculations, the viewing angles estimated with our new method are $1.8452 \times 10^{-9} \%$ and $2.4658 \times 10^{-9} \%$ for M87 and ${\rm Sgr\; A^\ast}$ small, respectively, compared to those estimated with the existing method. Therefore, we can say that the finite-distance effect is negligible for observing M87 and ${\rm Sgr\; A^\ast}$ from the Earth. In other words, our analysis supports the methodology and results of Event Horizon Telescope Collaboration~\cite{EventHorizonTelescope:2019dse, EventHorizonTelescope:2019uob, EventHorizonTelescope:2019jan, EventHorizonTelescope:2019ths, EventHorizonTelescope:2019ggy, EventHorizonTelescope:2021bee, EventHorizonTelescope:2021srq, EventHorizonTelescope:2022xnr, EventHorizonTelescope:2022vjs, EventHorizonTelescope:2022wok, EventHorizonTelescope:2022exc, EventHorizonTelescope:2022urf, EventHorizonTelescope:2022xqj}.

We explain that our results cannot be derived from those in the past papers analyzing the black hole shadows such as Refs.~\cite{Chael:2021rjo, Dokuchaev:2020wqk}. 

Let us clarify in what situation our method presented in this paper is useful, while we have already seen that our method makes no significant difference from the existing methods such as those in Refs.~\cite{Chael:2021rjo, Dokuchaev:2020wqk} for M87 and Sgr A*. The situation in which our method is essential is when $r_o/M$ is not known in advance of the observation of shadow. In such a situation, the validity of existing methods adopting the Bardeen coordinates and assuming sufficiently large $r_o/M$ cannot be examined in advance. Therefore, our method, which is valid for any values of $r_o/M$, should be adopted.

Let us stress again that this paper contains some new results which cannot be obtained until one regards $r_o/M$ as a free parameter. For example, the shape of shadow in the analysis adopting the Bardeen coordinates (like those in Refs.~\cite{Chael:2021rjo} and~\cite{Dokuchaev:2020wqk}) does not change as $r_o/M$ changes, provided the inclination angle $i$ is fixed. On the other hand, as shown in this paper, not only the size but also shape of shadow indeed changes as $r_o/M$ changes, even when the inclination angle $i$ is fixed.

The effects of observer’s motion on the image should be taken into account when the black hole is assumed to be at finite distance and especially the black hole is rotating itself. The so-called zero-angular-momentum observers~\cite{Bardeen:1973xx} and Carter's observers~\cite{Grenzebach:2014fha} would be the candidates of appropriate observers when one calculates the shadows of rotating black holes. While we will consider such moving observers in our next paper considering the Kerr black holes, we do not pay special attention to the effects resulting from the motion of observers in this paper because the above two kinds of observers in Schwarzschild spacetime reduce to the observer supposed in the present paper.

In summary, our method is a new type of parameter-determination method. It needs neither the small-angle approximation nor the prior information about the mass or distance. It is how to exactly determine $(M/r_o,i)$ from the image alone. It will be indispensable in the future when one has the chance to observe a black hole sufficiently close to the Earth or an observer in space.

%-------------------------------------------%
\section{Conclusion}
\label{sec:conclusion}
%-------------------------------------------%

We have analyzed the shadow of the Schwarzschild black hole with a thin accretion disk. In the analysis, the mass of black hole $M$ and inclination angle $i$ are assumed to be unknown, and the distance from the black hole to the observer $r_o$ is assumed to be finite and unknown. The dependence of black hole shadow shape and flux on mass $M$, distance $r_o$, and inclination angle $i$ was investigated.

It was found that two black hole shadows are congruent if $r_o/M$ and $i$ are common between the two systems. However, even though the shadows are congruent, the information about the flux can be used to distinguish two such systems. Any two systems can be distinguished from each other if both the shadow shape and flux information are observationally obtained, under the assumption one knows mass accretion rate $\dot{M}$ independently.

In this paper, we have considered the Schwarzschild black hole surrounded by thin accretion disk for simplicity. The next step would be to consider more general black holes such as the Kerr black hole and more realistic models of accretion disks. The reversibility of the map from the parameter space to the image library should be examined in such realistic models. If the image library is extended to include various black hole solutions as models, and if a shadow image not included in such a image library is actually observed, it suggests the existence of a new black hole solution or the modification of gravitational theory. 

Finally, it would be challenging but interesting to construct a movie library involving the time-varying shadow images, as the preparation for future observations.

%-------------------------------------------%
\section*{Acknowledgements}
%-------------------------------------------%
U.M.\ would like to thank H.\ Sotani for useful discussions. Works of U.M.\ are partially supported by JSPS Kakenhi Grants Numbers JP18K03652, JP22K03623 and President Project (Creative Research) at Akita Prefectural University.

\appendix

%-------------------------------------------%

%-------------------------------------------%

\end{document}